\documentclass[fleqn,usenatbib]{mnras}

\usepackage{newtxtext,newtxmath}

\usepackage[T1]{fontenc}

\DeclareRobustCommand{\VAN}[3]{#2}
\let\VANthebibliography\thebibliography
\def\thebibliography{\DeclareRobustCommand{\VAN}[3]{##3}\VANthebibliography}

\usepackage{graphicx}	
\usepackage{amsmath}	

\title[Colour gradients \& $C_{A}$ for PSBs]{Relation between colour gradient and central asymmetric features for post-starburst galaxies at $z\sim0.8$}

\author[J.Fujimoto \& M. Kajisawa]{
Junya Fujimoto,$^{1}$
and Masaru Kajisawa,$^{1,2}$\thanks{E-mail: kajisawa@cosmos.phys.sci.ehime-u.ac.jp}
\\
$^{1}$Graduate School of Science and Engineering, Ehime University, Bunkyo-cho, Matsuyama 790-8577, Japan\\
$^{2}$Research Center for Space and Cosmic Evolution, Ehime University
}

\date{}

\pubyear{}

\begin{document}
\label{firstpage}
\pagerange{\pageref{firstpage}--\pageref{lastpage}}
\maketitle

\begin{abstract}

We investigated colour gradients of photometrically selected
post-starburst galaxies (PSBs) 
at $0.7 <z< 0.9$ in the COSMOS field as a function of central concentration
of asymmetric features, $C_{A}$, in order to understand
their origins. 
We measured the colour gradients for 33 PSBs, 332 quiescent galaxies (QGs),
and 1136 star-forming galaxies (SFGs) by using COSMOS $HST$/ACS
$I_{\rm F814W}$-band and COSMOS-DASH $HST$/WFC3 $H_{\rm F160W}$-band data.
We found that the colour gradient,
$\Delta (I-H) = (I-H)_{\rm in} - (I-H)_{\rm out}$, decreases with increasing
$C_{A}$ for all the three populations.
Only PSBs with $\log{C_{A}}>0.6$ show positive gradients, which suggests
that their central asymmetric features are caused by disturbed distribution of
relatively young stars near the centre.
The colour gradients are also closely related with half-light radius rather than stellar mass for all the populations.
The positive colour gradients and very small sizes of those PSBs with high $C_A$ suggest that a nuclear starburst caused by gas-rich major merger occurred in the recent past. 
On the other hand, similarly massive PSBs with $\log C_A < 0.6$ show the negative colour gradients, 
heavier dust extinction, and larger sizes, and their origins may be different from those PSBs with high $C_A$.

\end{abstract}




\section{Introduction}

In general, galaxies can be divided into two populations: star-forming galaxies (hereafter SFGs) and quiescent galaxies 
with little star formation (hereafter QGs).
At $z\lesssim 1$, these  two populations show different morphological properties; QGs tend to show centrally 
concentrated spheroidal shapes with little disturbed feature, while many SFGs have a disc with spiral patterns 
and a spheroidal bulge (e.g.,  \citealp{rob94}; \citealp{blu19}).
While the stellar mass density and number density of SFGs have remained
nearly constant at $z\lesssim 1$, those of QGs have increased over the time,
which suggest that some fraction of SFGs stop forming new stars by some 
mechanism(s) and evolve into QGs (\citealp{fab07}; \citealp{pen10}).
This transition from SFG to QG is one of the most important processes in galaxy evolution.
While many mechanisms to quench star formation on various time scales have
been proposed (e.g., \citealp{dek86}; \citealp{bar96}; \citealp{aba99}; \citealp{bir03}; \citealp{mar09}; \citealp{fab12}; \citealp{spi22}), it is still unclear which mechanism plays dominant role in galaxy evolution and how it depends on conditions such as environment, epoch, galaxy properties, and so on. 

Investigating the properties of galaxies in the transition phase is one of the powerful ways to reveal 
the physical mechanisms of quenching.
In this context, post-starburst galaxies (hereafter PSBs) that experienced a strong starburst followed by 
quenching in the recent past have been considered to be an important population and studied 
intensively (see \citealp{fre21}, for recent review).
PSBs are selected by strong Balmer absorption lines and weak/no nebular emission lines such as H$\alpha$ or [O{\footnotesize II}], which indicate a large contribution from A-type stars and
little star formation, respectively
(e.g., \citealp{zab96}; \citealp{dre99}; \citealp{qui04}).
Recently, photometric selection methods for PSBs with several colours or  
 SED fitting have been also proposed (e.g., \citealp{wil14}; \citealp{wil16}; \citealp{him23}).

Morphological features of PSBs can provide important clues to understand the physical causes of starbursts and rapid quenching of star formation.
Many previous studies investigated morphological properties of PSBs, and found that a significant fraction of PSBs at low and intermediate redshifts show asymmetric/disturbed features such as tidal tails (e.g., \citealp{zab96}; \citealp{bla04}; \citealp{tra04}; \citealp{yam05}; \citealp{yan08}; \citealp{pra09}; \citealp{won12}; \citealp{deu20}; \citealp{wil22}).
These results indicate that galaxy merger/interaction may be involved in the stage where PSBs are formed.
\citet{him23} measured central concentration of asymmetric features,
$C_{A}$ for PSBs at $z \sim 0.8$, and found that a significant fraction of
those PSBs show high $C_{A}$ values, while they looks like early-type
morphologies with relatively high concentration and low asymmetry except
for significant central asymmetric features.
Theoretical studies with numerical simulations predicted that a gas-rich major merger causes a disturbance of morphology 
and an inflow of gas into the centre, and then a strong starburst occurs in the central region 
(e.g., \citealp{bar91}; \citealp{bar96}; \citealp{bek01}).
The central asymmetric features seen in PSBs may indicate that such nuclear
starbursts are closely related with the rapid quenching of star formation followed by the PSB phase.

The colour or specific star formation rate (SSFR) radial gradients
measured with spatially resolved multi-band/spectral data are powerful
tools to investigate how the quenching of star formation proceeded within
the galaxy. 
\citet{mor19} investigated SSFR gradients for galaxies at $0.2<z<1.2$
in the GOODS fields as a function of position in the SFR-$M_{\rm star}$
diagram relative to the main sequence (MS) of SFGs.
They found that galaxies below MS show positive SSFR gradients with 
suppressed SSFR at the centre, 
while SFGs above MS show enhanced SSFRs in inner regions, i.e.,
negative SSFR gradients.
Several studies also found the similar positive SSFR gradients for 
SFGs with declining star formation activities at $z \sim $ 1--2
(\citealp{tac18}; \citealp{abd18}; \citealp{nel16}; \citealp{nel21}).
On the other hand, \citet{che24} studied various types of PSBs
at $z<0.15$ from MaNGA survey, and found that those galaxies with central
or ring-like PSB features show positive and negative gradients in
$D_{n}(4000)$ and EW(H$\delta_{A}$), respectively, which suggest 
that active star formation occurs in inner regions in the recent past.
\citet{li23} divided local PSBs from the MaNGA survey into post-mergers
and non-mergers, and found that post-merger PSBs preferentially show
the outside-in quenching, while non-merger PSBs show both inside-out and
outside-in quenching equally.
\citet{deu20} also reported that PSBs at $z \sim 0.8$ have negative
gradients of H$\delta_{A}$.
Some of the theoretical studies mentioned above found that gas-rich major merger simulations show the positive colour (negative H$\delta$) gradients in the remnants (e.g., \citealp{bek05}; \citealp{sny11}; \citealp{zhe20}).
If such gas-rich mergers are closely related with the origins of PSBs, correlation between the colour gradients and morphological features such as $C_A$ could be seen in PSBs.

In this paper, we investigated colour gradients of PSBs at $0.7<z<0.9$ as
a function of morphological properties, in particular, the concentration of asymmetric features, $C_{A}$,
and compared them with SFGs and QGs in order to reveal the physical origins of
those PSBs and their quenching processes.
We measured the colour gradients of galaxies over a 0.66 deg$^2$ area in the
COSMOS field with 
$HST/{\rm ACS}$ data from COSMOS \citep{sco07}
and $HST/{\rm WFC3}$ data obtained by COSMOS-DASH \citep{mow19}.
Section 2 describes the data used in this study.
We describe sample selection in Section 3 and 
methods to measure the colour gradients in Section 4.
In Section 5, we present the colour gradients of the sample galaxies
and their relation with the morphological properties.
We discuss our results and their implications in Section 6 
and summarise the results of this study in Section 7.
Throughout this paper, we assume a flat universe with $\Omega_{\rm m}=0.3$, $\Omega_{\Lambda}=0.7$, 
and $H_0=70 {\rm \ km\ s^{-1}\ Mpc^{-1}}$, and magnitudes are given in the AB system.

\section{Data} \label{sec:data}

We used COSMOS $HST/{\rm ACS}$ $I_{\rm F814W}$-band data version 2.0 
(\citealp{koe07}) and $HST/{\rm WFC3}$ $H_{\rm F160W}$-band 
data version 1.2.10 from COSMOS-DASH survey (\citealp{mow19}) in this study.
The ACS $I_{\rm F814W}$-band data have a pixel scale of 0.03 arcsec/pixel and 
a PSF FWHM of $\sim$ 0.1 arcsec.
The WFC3 $H_{\rm F160W}$-band data have a pixel scale of 0.1 arcsec/pixel and 
a PSF FWHM of 0.21 arcsec (\citealp{mom17}).
The observation areas are 1.64 $\mathrm{deg^2}$ for the $I_{\rm F814W}$-band data
and 0.66 $\mathrm{deg^2}$ 
(including archival data) for the $H_{\rm F160W}$-band data.
The $5\sigma$ limiting magnitudes are $I \sim27.2$ mag for a 0".24 aperture
and $H \sim25.1$ 
mag for a 0".3 aperture.

\section{Sample}\label{sec:sample}
\subsection{Sample selection}\label{sec:samsel}
In this study, we used the same sample of PSB galaxies at $0.7<z<0.9$
in the COSMOS field as used in \citet{him23}.
We here briefly summarise the selection method, and refer the reader to
\citet{him23} for further details.

In order to select those galaxies that experienced a starburst followed
by rapid quenching several hundreds Myr before observations,
\citet{him23} fitted multi-band photometry of objects with $i<24$
from COSMOS2020 catalogue \citep{wea22} with population synthesis
models of GALAXEV \citep{bru03}.
X-ray AGNs were excluded from the sample, because model templates used 
in the SED fitting do not include AGN emission.
They used non-parametric, piece-wise constant function of star formation
history (SFH).
The look-back time for each galaxy was divided into seven periods,
namely, 0--40 Myr, 40--321 Myr, 321--1000 
Myr, 1--2 Gyr, 2--4 Gyr, 4--8 Gyr, and 8--12 Gyr before observation.
Model SED templates of stars formed in the different periods were
constructed assuming Chabrier initial 
mass function (\citealp{cha03}) and constant SFR in each period.
The model SEDs used in the fitting were based on a linear combination of
the seven templates, and normalisation 
coefficients for the seven templates were free parameters.
Those templates whose minimum ages are larger than the age of the universe
at the redshift were excluded from the fitting.
When the age of the universe entered between the minimum and maximum ages
of a template, we replaced it with a new template with the maximum age
slightly younger than the age of the universe.
In order to search for the best-fitting values of the normalisation
coefficients that provide the minimum $\chi ^2$, they 
adopted Non-Negative Least Squares (NNLS) algorithm \citep{law74} following
GASPEX by \citet{mag15}, while 
they used a simple full grid search for redshift, metallicity, and dust extinction.
The templates with three stellar metallicities, namely, $0.2$, $0.4$, and $1.0$
$Z_{\odot}$ were fitted.
For simplicity, they fixed the metallicity over all the periods except for the youngest period, 0--40 Myr before observation.
They added nebular emission only in the youngest template using PANHIT (\citealp{maw16}; \citealp{maw20}) because a contribution from the nebular emission is negligible in templates of the other older periods.
For the dust extinction, they used the Calzetti law \citep{cal00} and attenuation curves 
for local star-forming galaxies with different stellar masses, namely $10^{8.5}-10^{9.5}  M_{\odot}$, 
$10^{9.5}-10^{10.5}  M_{\odot}$,  and 
$10^{10.5}-10^{11.5}  M_{\odot}$, from \citet{sal18}.
Different ranges of $E(B-V)$ are adopted 
for the different attenuation curves, 
namely, $E(B-V) \leq 1.6$ for 
the Calzetti law and $E(B-V) \leq 0.4$ for those from \citet{sal18}.
They carried out Monte Carlo simulations to estimate probability distributions of derived physical properties such as stellar mass 
at the observed epoch, SFR, and specific star formation rate (SSFR) in the periods of look-back time.
They did 1000 simulations for each object, where the multi-band fluxes were fluctuated according to measurement errors and the same SED fitting with the simulated fluxes were performed, and adopted the median values as the physical properties.

They selected PSBs that experienced active star formation followed by rapid quenching several hundreds 
Myr before observation, by using SSFRs in the three youngest periods of look-back time, namely, 
${\rm SSFR_{0-40 Myr}}$, ${\rm SSFR_{40-321 Myr}}$, and ${\rm SSFR_{321-1000 Myr}}$.
Note that these SSFRs are defined as SFRs in these periods divided by stellar mass at the observed epoch (e.g., 
${\rm SSFR_{0-40Myr}=SFR_{0-40Myr}}/M_{\rm star,0}$) to easily compare the SSFRs among different periods.
Their PSB selection criteria are
\begin{equation}
  \begin{split}
  &{\rm SSFR}_{\rm 321-1000Myr} > 10^{-9.5} \hspace{1mm} {\rm yr}^{-1} \hspace{1mm} \&\\   
  &{\rm SSFR}_{\rm 40-321Myr} < 10^{-10.5} \hspace{1mm} {\rm yr}^{-1} \hspace{1mm} \&\\ 
  &{\rm SSFR}_{\rm 0-40Myr} < 10^{-10.5} \hspace{1mm} {\rm yr}^{-1}.
  \end{split}
  \label{eq:psbsel}
  \end{equation}
Since the distributions of SSFR in the periods of 0--40 Myr, 40--321 Myr, and 321--1000 Myr for galaxies with $i<24$ 
at $0.7<z<0.9$ show a peak at $\sim 10^{-9.5}$ yr$^{-1}$,
this selection picked up those galaxies 
whose SSFRs were comparable to or higher than the main sequence of SFGs in 321--1000 Myr
before observation and then decreased at least by an order of magnitude in the last 321 Myr.
They excluded galaxies with the reduced minimum $\chi^2>5$ and galaxies with nearby bright sources 
that affect the objects in the $HST/{\rm ACS}$ $I_{\rm F814W}$-band images from the sample.
Finally, there are 17459 galaxies with $i<24$ and reduced $\chi^2<5$ at $0.7<z<0.9$ in the COSMOS ACS field, and they selected 94 PSBs.

They constructed comparison samples of normal SFGs and QGs in the same redshift range.
For SFGs, they selected galaxies on and around the main sequence at least in the last $\sim 300$ Myr, namely, 
SSFR$_{0-40{\rm Myr}}=10^{-10}-10^{-9} {\rm yr^{-1}}$ and SSFR$_{40-321{\rm Myr}}=10^{-10}-10^{-9} {\rm yr^{-1}}$.
For QGs, they selected galaxies with low SSFRs within recent 1 Gyr, namely, 
SSFR$_{0-40{\rm Myr}}<10^{-10.5}{\rm yr^{-1}}$, 
SSFR$_{40-321{\rm Myr}}<10^{-10.5}{\rm yr^{-1}}$, 
and SSFR$_{321-1000{\rm Myr}}<10^{-10.5}{\rm yr^{-1}}$.
They selected 6581 SFGs and 670 QGs with $i<24$ and the reduced $\chi^2<5$ at $0.7<z<0.9$.

In this study, we used PSBs, QGs, and SFGs with $\log M_{\rm star} /M_{\odot} > 10$
for which both the $HST/{\rm ACS}$ 
$I_{\rm F814W}$-band data and $HST/{\rm WFC3}$ $H_{\rm F160W}$-band data  
are available in order to investigate their radial colour gradients.
We set the stellar mass limit to secure the completeness for PSBs with $i<24$
at $0.7<z<0.9$ \citep{him23}, while the magnitude limit of $i<24$ ensures
 accuracy of the colour gradient and morphological index described below.
Finally, we selected 33 PSBs, 332 QGs, and 1136 SFGs with $i<24$, $\log M_{\rm star} /M_{\odot} > 10$, and the reduced $\chi^{2} < 5$ at $0.7<z<0.9$.

For each PSB galaxy, we used SSFR values in 1000 Monte Carlo simulations to estimate a probability that the PSB criteria are satisfied (hereafter, PSB probability). In Appendix A, we examine physical properties of PSBs as a function of this probability to investigate how possible contaminants from non-PSB galaxies could affect our results.
We also use the PSB probability as a weight to estimate mean radial colour profiles of PSBs.

\begin{figure*} 
  \includegraphics[width=2\columnwidth]{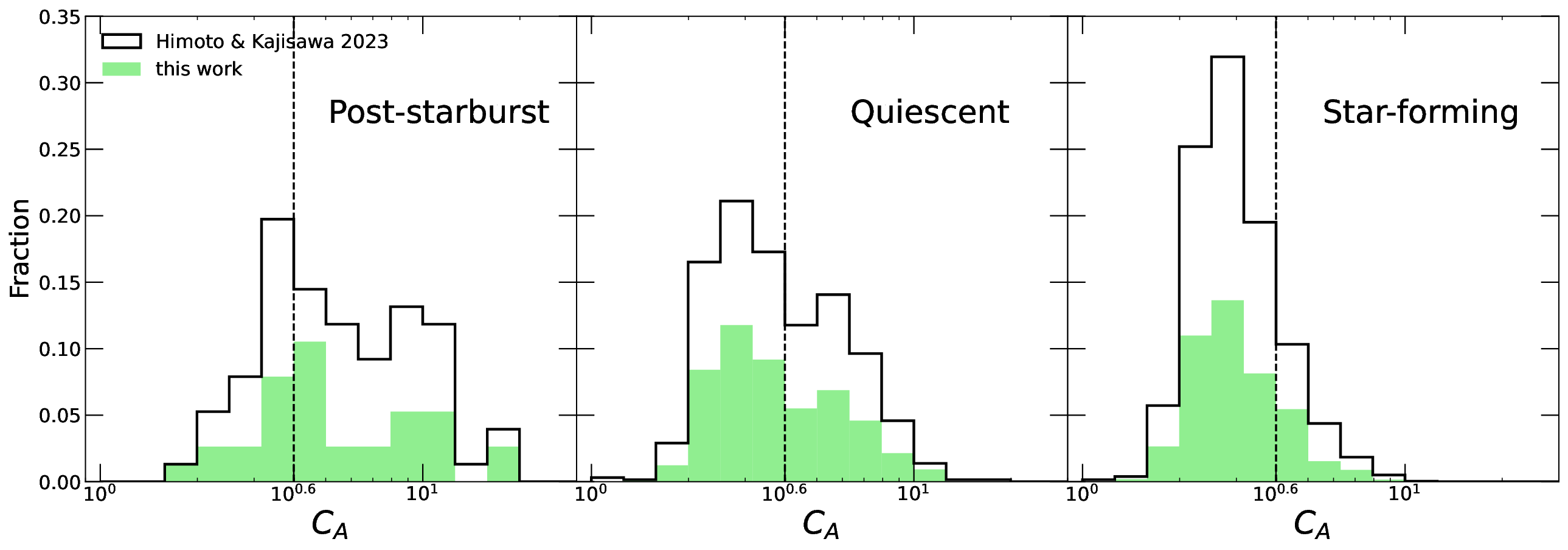}
  \caption{ 
  Distribution of the central concentration of asymmetric features, $C_{A}$ for PSBs (left), QGs (middle), and SFGs (right)   
  in \citealp{him23} (black solid) and in this study (light-green solid).
  Those galaxies from \citet{him23} are also limited to
  $M_{\rm star} > 10^{10} M_{\odot}$ for comparison.
  \label{fig:cahi}}
\end{figure*}

\subsection{Concentration of asymmetric features $C_A$}
We used a morphological index measuring central concentration of asymmetric features of the object, $C_A$,
which was newly devised and measured in \citet{him23}.
\citet{him23} devised $C_{A}$ to differentiate asymmetric features such as central disturbances or tidal tails.
They measured this index on the $HST$/ACS $I_{\rm F814W}$-band images.
At first, they rotated the object image around the object centre 
by 180 deg and subtracted the rotated image from the original image. 
Second, they calculated $C_A=r_{A,80}/r_{A,20}$, where $r_{A,80}$ and $r_{A,20}$ are radii which contain 80 per cent and 20 per cent of total flux of the rotation-subtracted image.
In measurements of $r_{A,80}$ and $r_{A,20}$, they corrected for 
contribution from the background noise 
with rotation-subtracted sky images, which were made from sky images with the 
similar procedure with the rotation-subtracted object image (\citealp{him23}
for details). 
They also repeated the calculations with 20 random sky images and
adopted the mean and standard deviation as $C_A$ and its uncertainty, respectively.

\begin{figure} 
  \includegraphics[width=1.02\columnwidth]{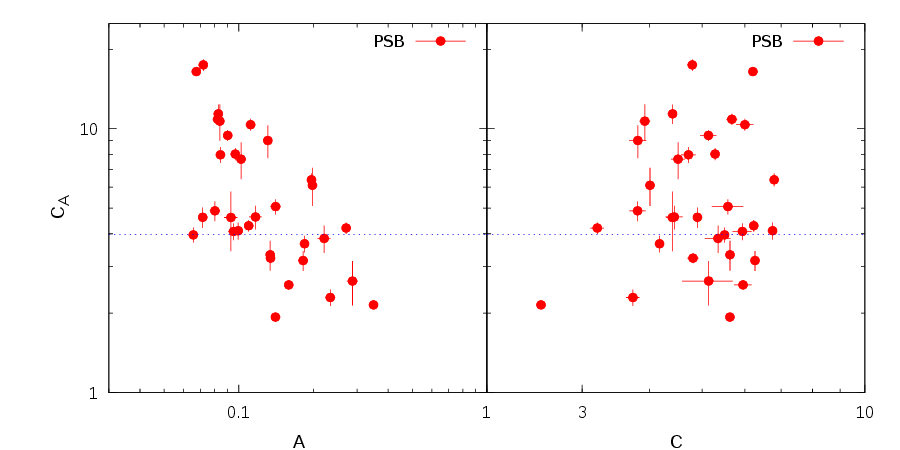}
  \caption{ 
    Distribution of $C_{A}$ for PSBs in this study as a function of
    Asymmetry index $A$ (left), and Concentration index $C$ (right).
    The dotted line shows the boundary of $\log{C_{A}} = 0.6$. 
  \label{fig:caac}}
\end{figure}

In Fig. \ref{fig:cahi}, we show distributions of $C_A$ for PSBs, QGs,
and SFGs in our sample (i.e., those with the both $I_{\rm F814W}$ and
$H_{F160W}$-band data), and those with $\log M_{\rm star} /M_{\odot} > 10$
from \citet{him23}.
The sample in this study seems to represent that of \citet{him23} over
the 1.64 deg$^{2}$ field in terms of $C_{A}$.
We divide those galaxies into those with high and low $C_{A}$ by
$\log{C_{A}} = 0.6$ in the following sections, and 
Table \ref{tab:sample_table} summarises the numbers of galaxies in our samples,
namely, PSBs, QGs, and SFGs with $\log C_{A} >0.6$ and $\log C_{A}<0.6$.
Fig. \ref{fig:caac} shows distribution of $C_{A}$ for PSBs in our sample
as a function of Asymmetry index $A$ and Concentration index $C$, which
were also estimated in \citet{him23}.
The $C_{A}$ values tend to decrease with increasing $A$, while there are
some PSBs with $\log C_{A} >0.6$ at $A \gtrsim 0.2$.
Those PSBs with $\log C_{A} > 0.6$ show relatively high $C$ values, while 
there are also many PSBs with $\log C_{A} < 0.6$ at $C \gtrsim 5$.
These trends are also seen in \citet{him23} (their Fig. 9).

\subsection{physical properties of the sample galaxies}
\begin{figure*} 
  \includegraphics[width=2\columnwidth]{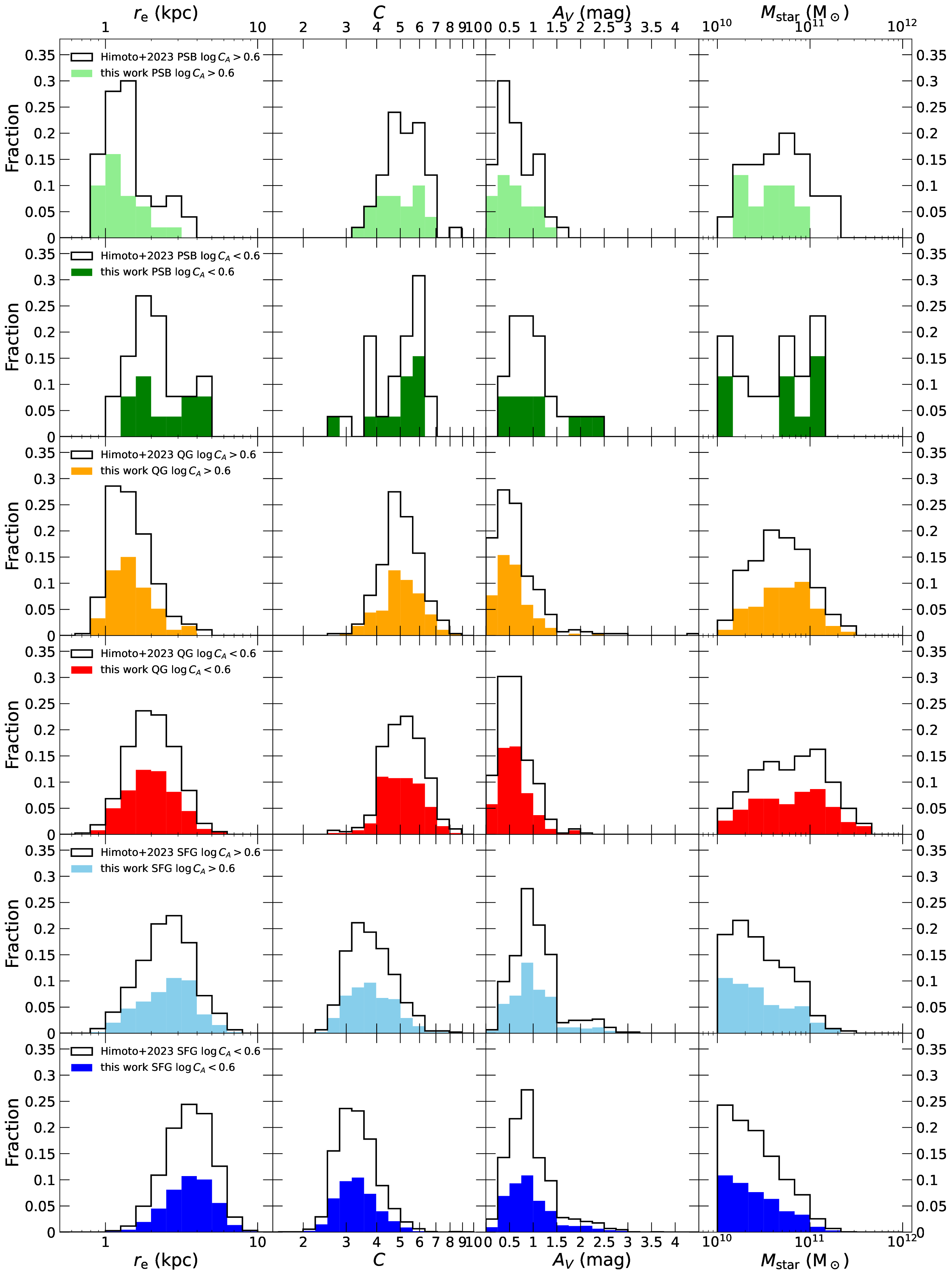}
  \caption{ 
    Distribution of $r_{\rm e}$, $C$, $A_V$, and $M_{\rm star}$ for PSBs with $\log C_A >0.6$ and 
    $\log C_A<0.6$ (top two panels), QGs with $\log C_A >0.6$ and $\log C_A<0.6$ (middle two panels), 
    and SFGs with $\log C_A >0.6$ and $\log C_A<0.6$ (bottom two panels)  
    in \citet{him23} (black solid) and in this study (colour solid).
    Those galaxies from \citet{him23} are also limited to
    $M_{\rm star} > 10^{10} M_{\odot}$ for comparison.
  \label{fig:size}}
\end{figure*}

In Fig. \ref{fig:size}, we show distribution of half-right radius $r_{\rm e}$, concentration $C$, dust extinction $A_V$, and stellar mass $M_{\rm star}$ for PSBs, QGs, and SFGs with high $C_A$ and low $C_A$ in our sample (i.e., those with the both $I_{\rm F814W}$ and
$H_{F160W}$-band data), and those with $\log M_{\rm star} /M_{\odot} > 10$ from \citet{him23}.
\citet{him23} measured the concentration index defined as $C=r_{80}/r_{20}$, where $r_{80}$ and $r_{20}$ are radii which contain 80 per cent and 20 per cent of the total flux of the object.
They measured a growth curve with circular apertures centred at a flux-weighted mean position of the object pixels to estimate $r_{80}$ and $r_{20}$.
In measurements of the growth curve, the subtraction of the background contribution was performed with the sky image.
They made 20 sky images for each object, repeated the measurements of $C$, and adopted their mean and standard deviation as $C$ and its uncertainty, respectively.
We measured half-light radii $r_{\rm e}$ of the all sample galaxies using the SExtractor software
version 2.5.0 (\citealp{ber96}) with the unconvolved $I_{F814W}$ band data.

\begin{table}
	\centering
	\caption{Sample sizes for PSBs, QGs, and SFGs with high and low $C_{A}$ values.}
	\label{tab:sample_table}
	\begin{tabular}{ccccc} 
		\hline
		Sample & total & $\log C_A > 0.6$ & $\log C_A  < 0.6$ \\
		\hline
		PSBs & 33 & 22 & 11 \\
		QGs & 332 & 131 & 201 \\
		SFGs & 1136 & 209 & 927 \\
	\end{tabular}
\end{table}

The distributions of these physical properties of our sample seem to represent those of the original sample.
For all three populations, those with high $C_{A}$ values have 
systematically smaller $r_{\rm e}$ than those with low $C_{A}$ in the population.
In particular, most of PSBs with $\log{C_{A}} > 0.6$ have $r_{\rm e} < 2$ kpc.
The stellar mass distributions for all PSBs, QGs, and SFGs do not strongly
depend on $C_{A}$.
While some PSBs with high $C_A$ in \citet{him23} have $M_{\rm star}> 10^{11}$
$M_{\odot}$, 
there are no PSBs with high $C_A$ values at $M_{\rm star} > 10^{11} M_{\odot}$ in our sample,  
simply because they are located outside the COSMOS-DASH survey area.

\section{Colour gradients} \label{sec:ana}

\subsection{PSF matching \& ZP correction}
We used the $I_{\rm F814W}$ and $H_{\rm F160W}$-band data to measure colour gradients
of the sample galaxies.
To measure the colour of the galaxies accurately as a function of radius, it is necessary to ensure that 
fluxes are measured in the same physical regions between the both bands.
Since the PSF sizes are different between the two bands, 
we matched the PSFs of the $I_{\rm F814W}$-band images
to those of the $H_{\rm F160W}$-band ones.
At first, we cut out a $12" \times 12"$ image 
for each sample galaxy in the both bands.
We also cut out bright point sources for each $10' \times 10'$ tile,
and stacked their images with a clipping mean algorithm in each band.
We then used IRAF/PSFMATCH task to create a kernel for the PSF matching
from the stacked images in the both bands and convolve the $I_{\rm F814W}$-band 
object images with the kernel for the corresponding tile.

\begin{figure*} 
  \includegraphics[width=2\columnwidth]{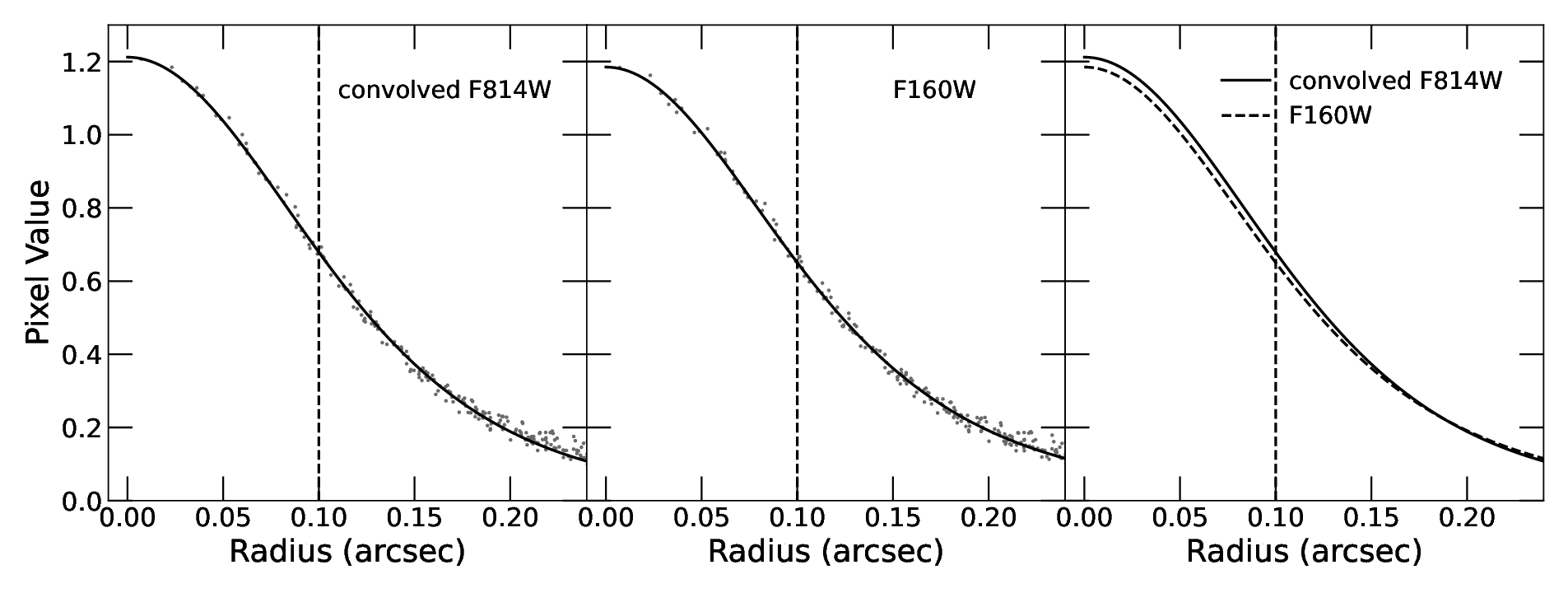}
  \caption{ Examples of the normalised PSF for matched $I_{\rm F814W}$-band (left) and $H_{\rm F160W}$-band data (middle).
  The lines represent the fitted Moffat function.
  The right panel compares the fitted Moffat profiles of the both bands. 
  The PSFs are normalised to the total counts of 100.
  \label{fig:psf}}
\end{figure*}

\begin{figure} 
  \includegraphics[width=0.97\columnwidth]{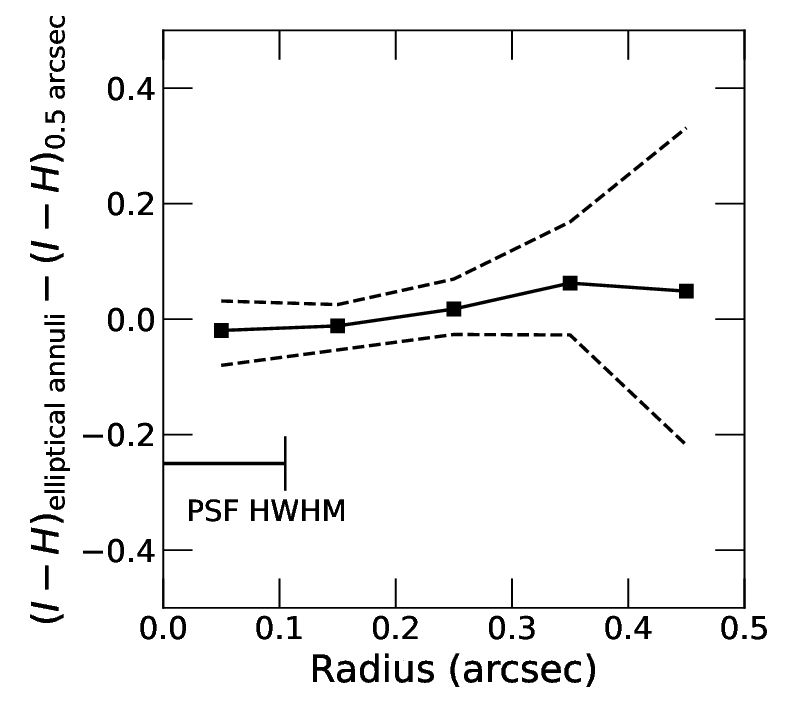}
  \caption{
    $I_{\rm F814W}-H_{\rm F160W}$ colour gradient for relatively bright
      isolated stars.
    Their colours are measured with a circular aperture with a 0.1'' radius
    and elliptical annuli whose inner and outer radii are 0.1'' and 0.2'', 
    0.2'' and 0.3'', 0.3'' and 0.4'', and 0.4'' and 0.5''.   
    All those colours in the different radii were converted to values 
    relative to the entire colour of the point source, which was
    measured with a circular aperture with 0.5'' radius. 
   Squares show the median relative colour values of the point sources, and the 
   dashed lines represent 16 and 84 percentiles.
  \label{fig:starih}}
\end{figure}

We checked PSFs in the $H_{\rm F160W}$-band and convolved $I_{\rm F814W}$-band images to examine their effects on the colour measurement.
In Fig. \ref{fig:psf}, we show examples of the PSF for the both bands.
The shapes of the normalised PSF for the two bands are similar, which suggests that even a 0.1 arcsec radius would not affect the colour measurement.
We measured the $I-H$ colour gradients of relatively bright and isolated
  stars with a central aperture and annuli of various sizes, and then subtracted
the colour measured within 0.5'' radius of each point source from the values
in all aperture and annuli.
Fig. \ref{fig:starih} shows distribution of the relative colour gradient
for those point sources.
The median values are consistent with no systematic offset at least at
$R < 0.3''$, while the measurement errors become large at outer annuli.
Thus the errors in the PSF matching do not seem to strongly affect the
colour gradients.

\begin{figure*} 
  \includegraphics[width=1.8\columnwidth]{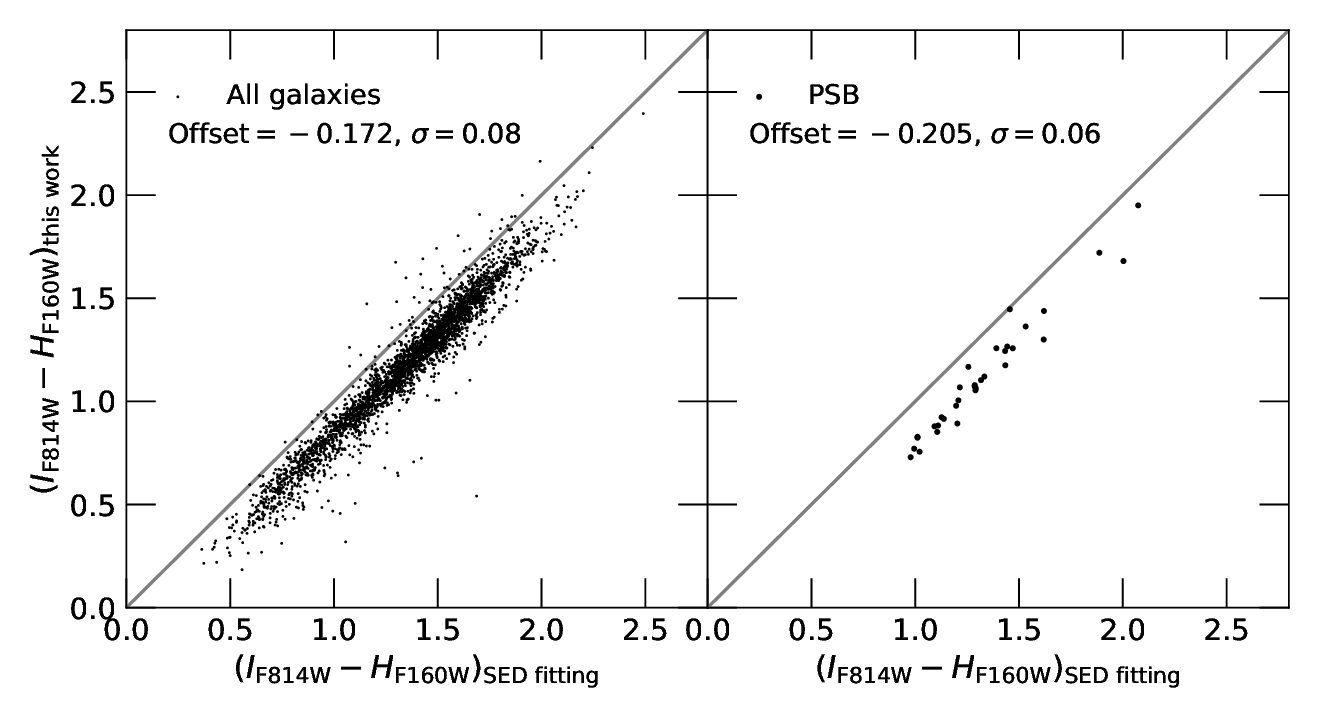}
  \caption{ Comparison between the $I_{\rm F814W}-H_{\rm F160W}$ colours measured on the HST data and those expected from the best-fit templates in the SED fitting from UV to NIR wavelength, for all galaxies with $M_{\rm star} > 10^{10} M_{\odot}$ (left panel) and PSBs (right panel).
  \label{fig:offset}}
\end{figure*}

We also checked if the measured $I_{\rm F814W}-H_{\rm F160W}$ colour over the galaxy is consistent with the photometric SED from the COSMOS2020 catalogue.
In Fig. \ref{fig:offset}, we compare the $I_{\rm F814W}-H_{\rm F160W}$ colours measured with an aperture of 2.5 times Kron radius with those expected from the best-fit templates in the SED fitting for all sample galaxies and PSBs.  
In all sample galaxies with $M_{\rm star} > 10^{10} M_{\odot}$, total $I-H$ colours are offset by $0.17\pm 0.08$ mag from those expected from the best-fit templates.
In PSBs, total $I-H$ colours are offset by $0.21\pm 0.06$ mag.
We adopted the value for all sample galaxies as zero-point offset, and corrected the $I_{\rm F814W}-H_{\rm F160W}$ colours measured in all apertures and annuli. 

\subsection{colour gradients}
We measured fluxes of the sample galaxies on the convolved $I_{\rm F814W}$-band 
images and the $H_{\rm F160W}$-band images using the SExtractor.
While the unconvolved $I_{\rm F814W}$-band images were used as the detection
images in the SExtractor runs, we masked the other objects in the images 
before the flux measurements by using the SEGMENTATION images,  
which were made by running SExtractor on the Subaru/Suprime-Cam
$i'$-band data and then aligned to the ACS $I_{\rm F814W}$-band images
(\citealp{him23} for details). 
In order to estimate colour gradients, we measured fluxes in 5 regions of the galaxy in the both bands.
The most inner region is a circular aperture with a 0.1 arcsec radius, which corresponds
to $\sim 0.75$ kpc for galaxies at $z \sim 0.8$.
For the other outer regions, we measured with elliptical annuli whose
inner and outer radii are 0.5 and 1.0, 1.0 and 1.5, 1.5 and 2.0, and 2.0 and 2.5 times
Kron radius of the galaxy..
The elliptical shape was calculated from the second-order moments of
the pixels belonging to the object by SExtractor.
The all regions were centred at a flux-weighted mean
position of the object pixels in the detection images.
Since the Kron radius was measured on the unconvolved $I_{\rm F814W}$-band
image, these radii are actually slightly smaller
than Kron radius for the convolved $I_{\rm F814}$-band and
the $H_{\rm F160W}$-band images.
We calculated colours in those regions as differences between
AB magnitudes in the both bands, namely, $I-H = m_{\rm F814W} - m_{\rm F160W}$.

\begin{figure*} 
  \includegraphics[width=1.8\columnwidth]{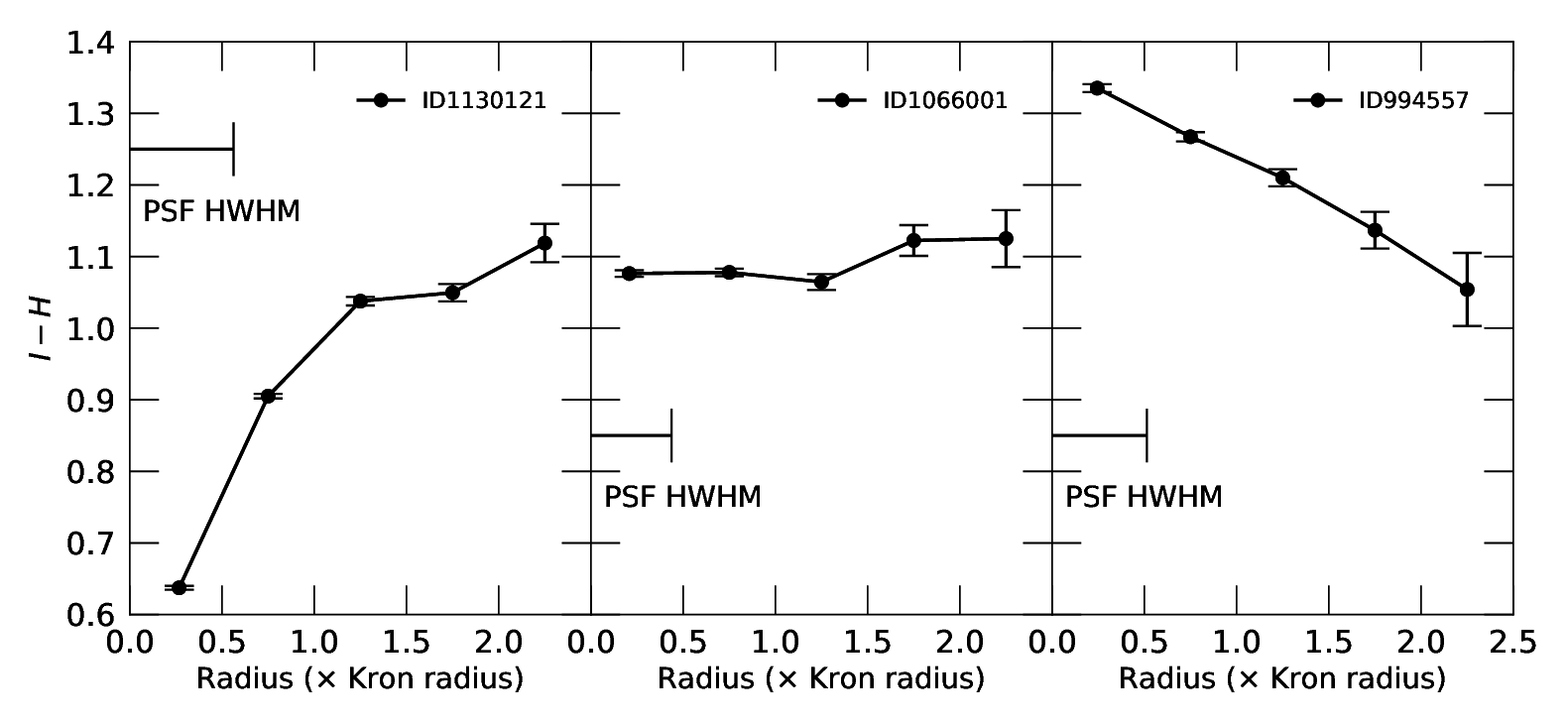}
  \caption{ Examples of the radial  $I-H$ colour profiles for PSBs with positive (left), relatively flat (middle), and negative (right) colour gradients.
    Half width at half maximum (HMHM) of PSF is also plotted in each panel.
  \label{fig:excg}}
\end{figure*}

In order to estimate uncertainty by the background fluctuation in the colour gradient,
we made sky images in the both bands 
by replacing pixels belonging to the object with randomly 
selected ones that do not belong to any objects
in a 30'' $\times$ 30'' field around the object.
The sky images for the $I_{\rm F814W}$-band data 
were convolved with the same kernel as for the object images.
We cut out 12'' $\times$ 12'' sky images at 200 random positions 
in the 30'' $\times$ 30'' field, and repeated the flux measurements 
on these images with the same manner to estimate background fluctuation.
We adopted their standard deviation as the flux error and calculated
the uncertainty in the colour gradients from the flux errors.
We show examples of the radial profiles of the $I-H$ colour in Fig. \ref{fig:excg}.

We quantify the colour gradients as difference between colours measured with the innermost aperture of 0.1 arcsec radius and the outermost elliptical annulus with inner and outer radii of 2.0 and 2.5 times Kron radius.
We defined the colour gradient as 
\begin{equation}
  \begin{split}
  &\Delta (I-H) = (I-H)_{\rm in} - (I-H)_{\rm out},
  \end{split}
  \label{eq:delih}
  \end{equation}
where $(I-H)_{\rm in}$ is $I-H$ colour in a circular aperture with a 0.1 arcsec radius, 
and  $(I-H)_{\rm out}$ is that in an elliptical annulus with inner and
outer radii of 2.0 and 2.5 times Kron radius of the object.
Thus, negative $\Delta(I-H)$ indicates a positive colour gradient.
Note that the distance between the central and outer regions is scaled
by Kron radius in this definition of the colour gradient.
In appendix \ref{sec:dlogr}, we also present results with $d(I-H)/d\log{r}$
rather than $\Delta(I-H)$.

We calculated mean radial profiles of the $I-H$ colour for the samples of
PSBs, QGs, and SFGs with $\log{C_{A}}<0.6$ and $\log{C_{A}}>0.6$.
Three types of the mean radial profiles of the $I-H$ colour were estimated, namely, the mean profile of the observed $I-H$ colour, 
that of the $I-H$ colour corrected for the dust extinction, and the weighted mean profile of the dust-corrected colour. 
The weighted mean profile was derived only for PSBs, and we used the PSB probability mentioned in Section \ref{sec:samsel} as a weight.
We corrected the colour in each radius for the dust extinction by assuming 
the constant $E(B-V)$ value estimated in the SED fitting
(Section \ref{sec:samsel}) over all the radii, and then 
calculated the mean of the corrected $I-H$ colour in each annulus
scaled with Kron radius.
We note that the corrected colours could be biased if the dust extinction systematically changes with radius.
For example, if central bulges in normal SFGs are less dusty than outer star-forming disks, the constant correction for dust extinction would underestimate and overestimate the colours in inner and outer regions, respectively.  
The dust extinction near the centre could also be enhanced, if a nuclear starburst occurred. 
We discuss possible effects of the dust extinction in Section \ref{sec:dis}.
We also estimated uncertainty of the mean radial profiles from the flux errors in the annuli for the sample galaxies.

\section{Results} \label{sec:res}
\subsection{Colour gradients}

\begin{figure*} 
  \includegraphics[width=2\columnwidth]{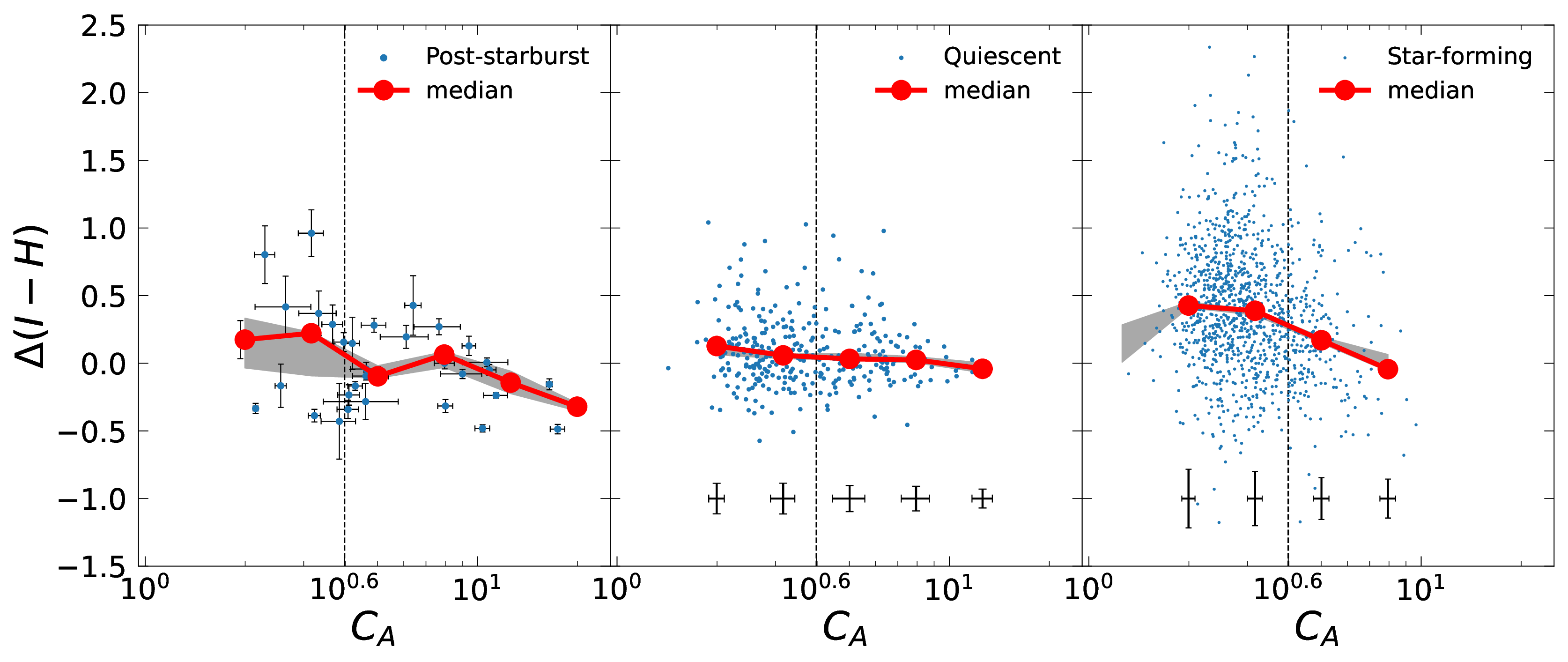}
  \caption{ 
  $\Delta(I-H) = (I-H)_{\rm in} - (I-H)_{\rm out}$ as a function of $C_A$ for PSBs (left), QGs (middle), and SFGs (right). 
  The inner and outer colours are measured with a circular aperture with 0.1 arcsec radius and an outer elliptical annulus with inner and outer radii of 2.0 and 2.5 times Kron radius of the object, respectively. 
  Circles show the median values of $\Delta(I-H)$ in $C_A$ bins with a width of $\pm 0.1$ dex, and the shaded regions represent 68\% confidence ranges of
  the median values.
  In the middle and right panels, the median values of the measurement errors
  in the $C_{A}$ bins are shown at bottom for clarity.
  \label{fig:deca}}
\end{figure*}

Fig. \ref{fig:deca} shows relationship between $C_A$ and $\Delta (I-H)$ 
for PSBs, QGs, and SFGs with $M_{\rm star} > 10^{10} M_{\odot}$ at 
$0.7 < z < 0.9$. The median $\Delta (I-H)$ as a function of $C_{A}$
  and its 68\% confidence range are plotted as circles and shaded regions
  in the figure.
  In order to estimate the 68\% confidence range of the median $\Delta (I-H)$,
  we carried out 1000 Monte Carlo simulations by adding random shifts based on
  the measurement errors to $\Delta (I-H)$ and $C_{A}$ of sample galaxies
  and calculating the median values in the $C_{A}$ bins.
One can see that galaxies with high $C_{A}$ values preferentially 
have relatively low $\Delta (I-H)$, while those with low $C_{A}$
show a wide range of $\Delta (I-H)$. 
Median values of $\Delta (I-H)$ tend to decrease with increasing
$C_A$ for all the three populations.
The median $\Delta (I-H)$ of SFGs with low $C_{A}$ is relatively high
($\sim 0.4$), and their median value decreases to $\sim 0$ at $C_{A} \sim 6$.
On the other hand, QGs tend to have flatter colour gradients, and their 
$\Delta (I-H)$ only weakly depends on $C_{A}$.
PSBs show the similar $C_{A}$ dependence of the colour gradient with SFGs,
but their median $\Delta (I-H)$ for a given $C_{A}$ is lower than SFGs.
Table \ref{tab:example_table} summarises the median values of $\Delta (I-H)$
for PSBs, QGs, and SFGs.
We divided our samples of the three populations into those
with $\log{C_{A}} < 0.6$ and $\log{C_{A}} > 0.6$, and the median values  
for these sub-samples with low and high $C_{A}$ are also shown in the table.
The errors in the table were estimated with the bootstrap resampling.
The median value of $\Delta (I-H)$ for PSBs is $\sim -0.05$, 
which is lower than those of QGs and SFGs ($\sim 0.05$ and $0.36$). 
This is mainly because PSBs have a higher fraction of those galaxies with
high $C_{A}$ values than QGs and SFGs (Fig. \ref{fig:cahi}). 
The median  $\Delta (I-H)$ for PSBs with high $C_{A}$ is the lowest
 value of $\sim -0.09$, 
which means that those galaxies show bluer colours in the central region.

\begin{table}
	\centering
	\caption{Median $\Delta (I-H)$ for the samples with the different SFHs and $C_{A}$. 
	The errors were estimated with the bootstrap resampling. \label{tab:median}}
	\label{tab:example_table}
	\begin{tabular}{cccc} 
		\hline
		 & all & $\log{C_{A}}>0.6$ & $\log{C_{A}}<0.6$ \\
		\hline
		PSBs & $-0.049 \pm 0.080$ & $-0.087 \pm 0.056$ & $0.175 \pm 0.201$ \\
                QGs & $0.050 \pm 0.013$ & $0.026 \pm 0.016$ & $0.074 \pm 0.015$ \\
                SFGs & $0.356 \pm 0.016$ & $0.139 \pm 0.032$ & $0.406 \pm 0.017$ \\
		\hline
	\end{tabular}
\end{table}

\begin{figure*} 
  \includegraphics[width=2\columnwidth]{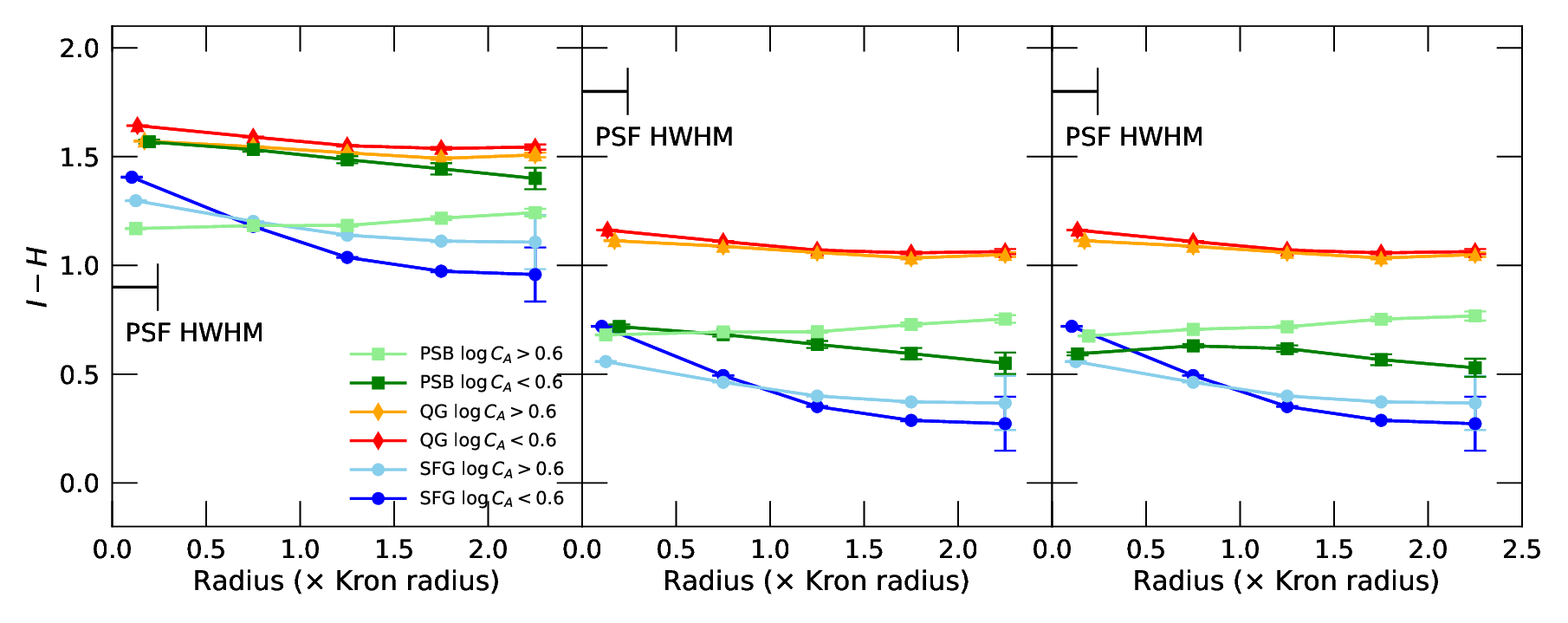}
  \caption{
    Mean radial $I-H$ colour profiles for PSBs (squares), QGs (diamonds), and SFGs (circles) 
    with $\log C_A <0.6$ (dark) and $\log C_A>0.6$ (light).
    $I-H$ colours are measured with a central circular aperture with 0.1'' radius and annuli with inner and outer radii 
    of 0.5--1.0, 1.0--1.5, 1.5--2.0, and 2.0--2.5 times Kron radius of the object.
    In the middle panel, these colours are corrected for dust extinction before the stacking (see text for details).
    The right panel is the same as the middle panel but the colours of PSBs with $\log C_A < 0.6$ and $\log C_A > 0.6$ are averaged by weighting each object by the PSB probability.
    The innermost colour is plotted at 0.05 arcsec on a scale of the Kron radius.
  \label{fig:cg}}
\end{figure*}

Fig. \ref{fig:cg} shows the mean radial profiles of the observed and dust-corrected $I-H$ colour 
for PSBs, QGs, and SFGs with $\log{C_{A}} < 0.6$ and $\log{C_{A}} > 0.6$.
Only PSBs with high $C_A$ values show a positive mean colour gradient,
which means the bluer colours at the inner radii, while the other samples 
show negative colour gradients.
In the observed colours, PSBs with low $C_A$ show significantly redder colours than those with high $C_A$, especially, at inner radii.
In the colours corrected for the dust extinction (middle panel), PSBs with low $C_{A}$ values show the similar but slightly redder $I-H$
in the central region, and their mean $I-H$ colour becomes bluer than
those with high $C_{A}$ at the outer radii.
The redder observed colours of PSBs with low $C_A$ are mainly caused by larger dust extinction than those with high $C_A$.  
SFGs with low $C_A$ values (normal SFGs) show the steepest negative colour
gradient where the colour decreases from the similar value with PSBs at the
centre to much bluer values in the outer regions.
In contrast, SFGs with high $C_A$ values have a flatter colour gradient,  
being bluer at the centre and redder at the outer radii than
SFGs with low $C_A$ values.
The mean colour profile of QGs does not strongly depend on $C_{A}$, and
they show slightly redder colours at the centre and relatively flat gradients
at the outer radii.

In the right panel of Fig. \ref{fig:cg}, we show the mean radial profiles of the dust-corrected colour weighted by the PSB probability for PSBs with low and high $C_A$ to examine possible effects of contamination from non-PSB galaxies.
In the panel, PSBs with high $C_A$ show the similarly positive colour gradient.
On the other hand, PSBs with low $C_A$ show bluer colour than those with high
$C_A$, even in the central region, and their colour gradient could be flatter.
Although the radial mean profile of those with low $C_{A}$ values could
  be affected by some contaminants, the difference between those with
  high $C_{A}$ and low $C_{A}$ still remains.

\begin{figure} 
  \includegraphics[width=1\columnwidth]{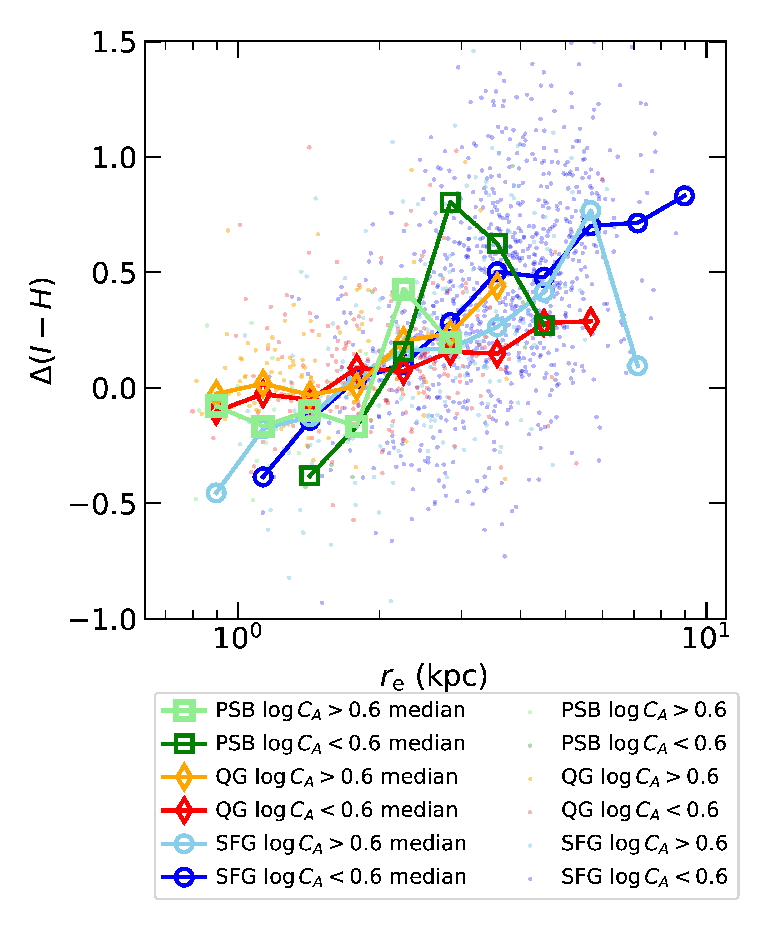}
  \caption{ 
  $\Delta (I-H)$ as a function of half-light radius, $r_{\rm e}$ for PSBs (squares), QGs (diamonds), and SFGs (circles) 
  with $\log C_A <0.6$ (dark) and $\log C_A>0.6$ (light).
  The solid lines show the median values of $\Delta (I-H)$ in $r_{\rm e}$
  bins with a width of $\pm 0.05$ dex.
  \label{fig:ihr}}
\end{figure}

In appendix \ref{sec:asycon}, we present $\Delta(I-H)$ as a function of
Asymmetry index $A$ and Concentration index $C$ for PSBs, QGs, and SFGs.
The $\Delta(I-H)$ distributions for the three populations do not significantly
depend on $C$. PSBs with relatively low $A$ values tend to have
$\Delta(I-H)\lesssim 0.3$, while those with high $A$ values show a wide range 
of $\Delta(I-H)$. Most of the PSBs with low $A$ values correspond to those 
with $\log{C_{A}}>0.6$ (Fig. \ref{fig:caac}).
$\Delta(I-H)$ of QGs does not strongly depend on $A$. The median $\Delta(I-H)$
of SFGs increases with increasing $A$, but scatters around the median values
are very large. Thus asymmetric features, in particular, their spatial
concentration are closely related with low $\Delta(I-H)$ values, i.e.,
flat or positive colour gradients.

\subsection{Size and mass dependence of the colour gradient}

In order to examine size dependence of the colour gradient, we show $\Delta(I-H)$ vs. $r_{\rm e}$ for PSBs, QGs, and SFGs with high and low $C_A$ values in Fig. \ref{fig:ihr}.
Those galaxies with smaller sizes tend to show lower $\Delta (I-H)$, 
and the median $\Delta (I-H)$ values decrease with decreasing $r_{\rm e}$ 
for all the samples.
The median $\Delta (I-H)$ for a given $r_{\rm e}$ does not significantly 
depend on $C_A$ for all the populations.
We confirmed that the size dependence of $\Delta (I-H)$ does not change
if we use half-light radii measured on the $H_{\rm F160W}$-band images
instead of the $I_{\rm F814W}$-band ones.
We also checked cases where we adopt $d(I-H)/d(\log{r})$ instead of
  $\Delta (I-H)$ in appendix \ref{sec:dlogr}, and found that  
  the dependence of the colour gradient both on $C_{A}$ and $r_{\rm e}$ seen in
  Figs. \ref{fig:deca} and \ref{fig:ihr} does not
  strongly change if we use $d(I-H)/d(\log{r})$.

\begin{figure} 
  \includegraphics[width=1\columnwidth]{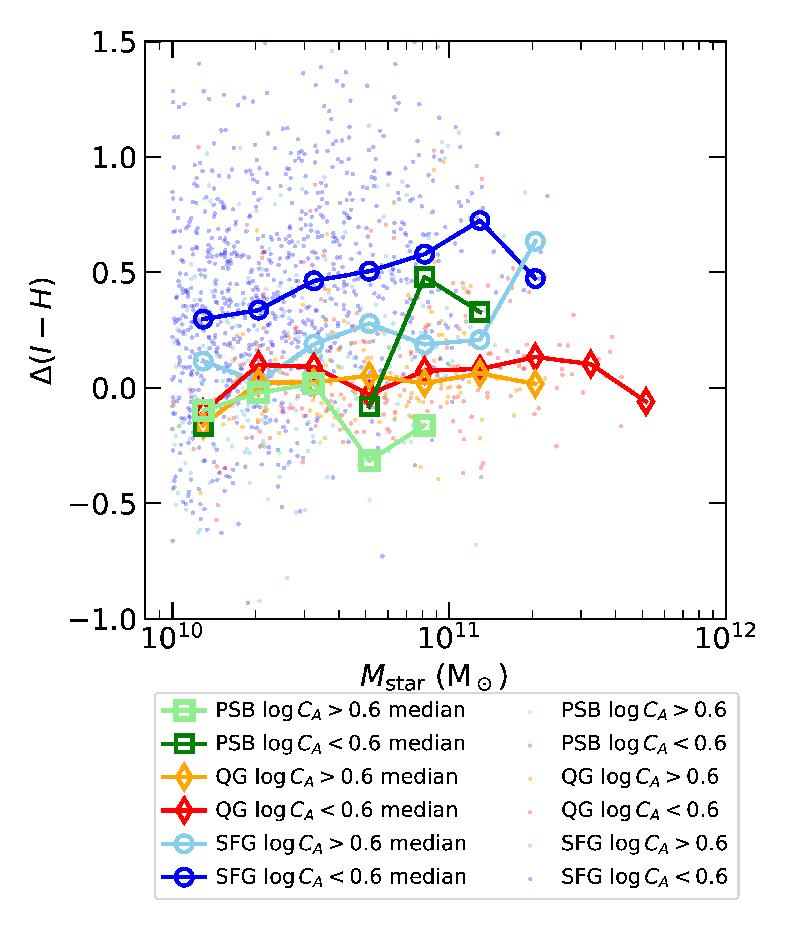}
  \caption{ 
  $\Delta (I-H)$ as a function of $M_{\rm star}$ for PSBs, QGs, and SFGs 
  with $\log C_A < 0.6$ and $\log C_A >0.6$. 
  The symbols are the same as Fig. \ref{fig:ihr}.
  The solid lines show the median values of $\Delta (I-H)$ in $M_{\rm star}$ bins with a width of $\pm 0.1$ dex.
  \label{fig:ihm}}
\end{figure}

In Fig. \ref{fig:ihm}, we show $\Delta (I-H)$ as a function of $M_{\rm star}$ 
for PSBs, QGs, and SFGs with high and low $C_A$ values.
Only SFGs with low $C_{A}$ values show a clear mass dependence of
the colour gradient, and more massive those galaxies tend to have steeper
negative gradients, which probably reflects the mass dependence of the bulge
fraction (e.g., \citealp{lan14}).
The other samples show no significant mass dependence of the colour gradient.
SFGs with $\log{C_{A}} > 0.6$ have much lower $\Delta (I-H)$ for a given
$M_{\rm star}$ than those with low $C_{A}$.
QGs with high $C_{A}$ seem to have slightly lower $\Delta (I-H)$ for a given
$M_{\rm star}$ than those with low $C_{A}$ over a wide range of $M_{\rm star}$.  
PSBs with $\log{C_{A}} > 0.6$ tend to have the lowest $\Delta (I-H)$
for a given $M_{\rm star}$, and their median $\Delta (I-H)$ is lower than
that of PSBs with low $C_{A}$ at $M_{\rm star} = $ $10^{10.6}$--$10^{11} M_{\odot}$, 
where the numbers of available galaxies in the both samples are sufficient
for the comparison.

\section{Discussion}\label{sec:dis}

In this study, we measured the $I-H$ colour gradients for PSBs at $z \sim 0.8$
in the COSMOS field 
using the ACS $I_{\rm F814W}$-band images and WFC3 $H_{\rm F160W}$-band images, 
and compared them with SFGs and QGs.
We found that only PSBs with high $C_A$ values show the positive gradients on average,  
which means that their central regions are bluer than outer regions. 
We here compare our results with previous studies and discuss implications
of the results for the evolution of these galaxies.

\cite{him23} found that a significant fraction of the PSBs at $z \sim 0.8$ have high $C_{A}$,  
and discussed that nuclear starbursts could cause disturbances in the young stellar and/or dust 
distribution near the centre, which leads to their asymmetric features in the central region.
We found that $\Delta (I-H)$ decreases with increasing $C_{A}$, and those PSBs with $\log{C_{A}} > 0.6$ 
preferentially show positive colour gradients.
Since the fraction of galaxies with high $C_A$ is relatively high in PSBs, the median colour gradient for all PSBs is positive in Table \ref{tab:example_table}.
\cite{yam05} measured $g-r$ colour gradients for 50 PSBs at $z<0.2$ from the SDSS survey, and found that these PSBs have a wide range of the colour gradients and 2/3 of them show positive gradients. 
\cite{yan08} investigated $B-R$ colour gradients for 21 PSBs at $z<0.2$ with HST data, and found similarly wide distribution of the colour gradients, where 12 PSBs have positive colour gradients and 5 those galaxies have negative ones. 
The colour gradients of PSBs and their $\Delta (I-H)$ distribution (Fig. \ref{fig:deca}) seem to be consistent with these previous studies for PSBs at $z<0.2$.

\begin{figure} 
  \includegraphics[width=1\columnwidth]{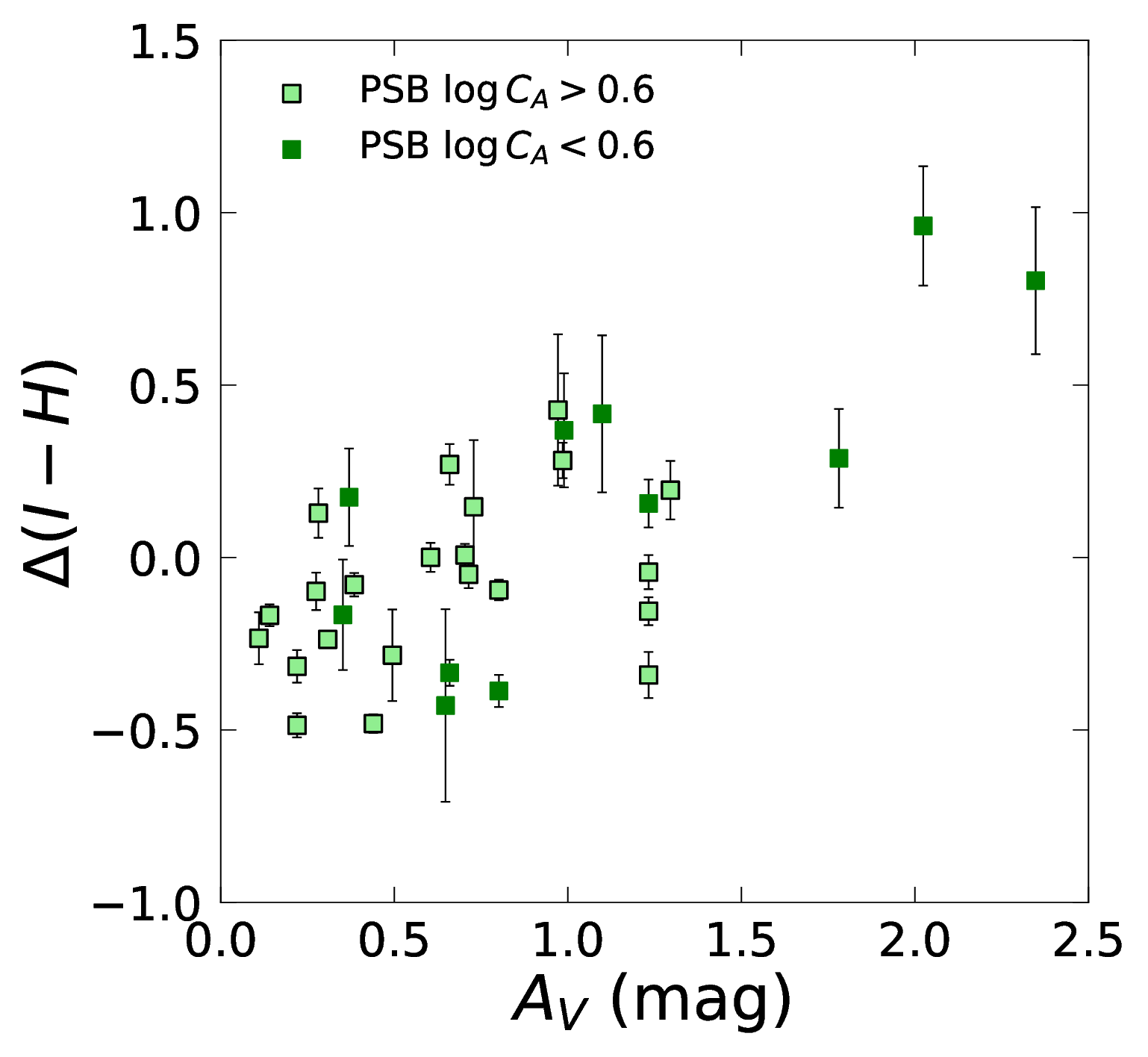}
  \caption{ $\Delta (I-H)$ as a function of $A_V$ for PSBs with $\log C_A>0.6$ (light green) and $\log C_A<0.6$ (dark green). 
  \label{fig:ihav}}
\end{figure}

The positive colour gradients for PSBs with $\log{C_{A}} > 0.6$ in Fig. \ref{fig:deca} suggest that  
the disturbed distribution of relatively young stellar population leads to their high $C_{A}$ and 
their central regions do not seem to be heavily obscured by dust.
Those PSBs with high $C_{A}$ have relatively small $A_{V}$ ($\sim 0.6$ mag)
as shown in Fig. \ref{fig:ihav}, and  
their mean corrected colour of $I-H \sim 0.5$ within central
$\lesssim 1$ kpc (Fig. \ref{fig:cg}) 
is consistent with stars formed in 321--1000 Myr before observation.  
We note that possible effects of the dust obscuration near the centre cannot be completely ruled out, 
since we examined only one colour that roughly corresponds to the rest-frame $B-I$ colour at $z \sim 0.8$.
The spatially resolved longer wavelength data by JWST enable us to resolve the age-dust degeneracy.

Previous studies with numerical simulations predicted that gas-rich major mergers cause gas infall to 
the centre of the remnants and then a strong nuclear starburst occurs, 
which leads to rapid quenching through gas consumption by the burst,
gas loss/heating by supernova/AGN feedback or tidal force, and so on
(e.g., \citealp{bar96}; \citealp{bek01}; \citealp{dav19}; \citealp{spi22}).
Some of such gas-rich merger simulations reproduced the positive colour
gradients in the PSB phase, probably because the gas infall to the centre
and/or gas loss from the galaxy by tidal force cause
a lack of cold gas in outer regions at earlier phase, and the starbursts 
in the inner regions tend to be stronger and continue to later times    
(\citealp{bek05}; \citealp{sny11}; \citealp{zhe20}).
Such gas-rich mergers followed by the nuclear starburst may explain that only PSBs with high $C_A$ show the positive colour gradients.

\begin{figure} 
  \includegraphics[width=1\columnwidth]{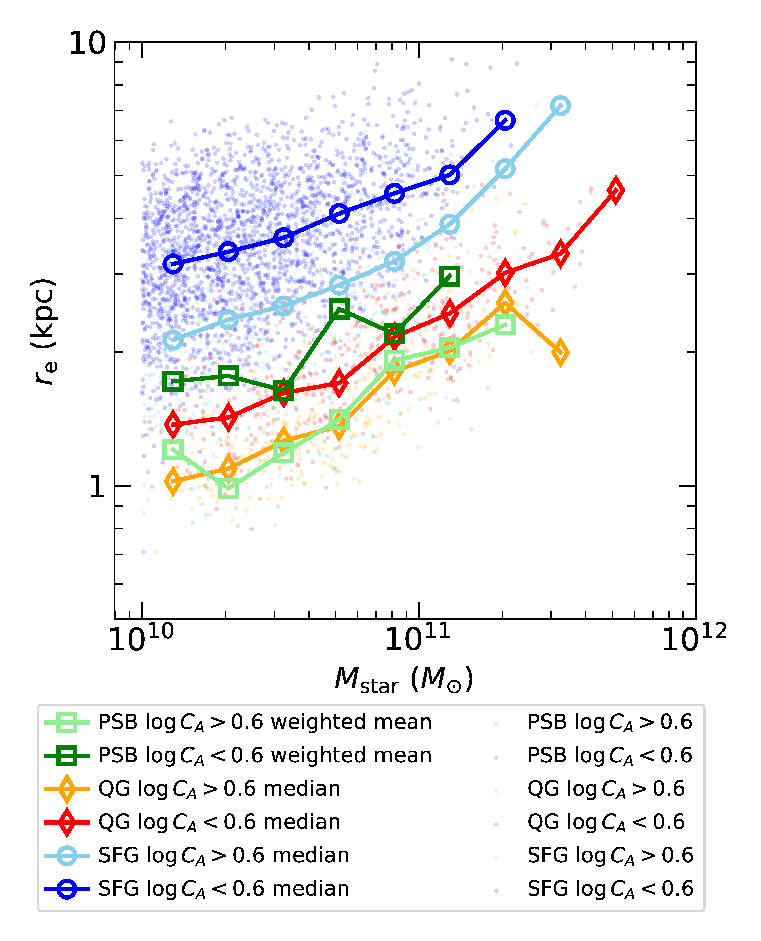}
  \caption{ 
  half-light radius $r_{\rm e}$ as a function of ${M_{\rm star}}$ for PSBs, QGs, and SFGs 
  with $\log C_A<0.6$ and $\log C_A>0.6$ in \citet{him23}.
  Those galaxies with $M_{\rm star} > 10^{10} M_\odot$ from \citet{him23} are plotted.
  The symbols are the same as Fig. \ref{fig:ihr}.
  The solid lines for QGs and SFGs show the median values of half-light radius in ${M_{\rm star}}$ bins with a width of $\pm 0.1$ dex.
  The solid lines for PSBs show the PSB probability weighted mean values in the same bins.
  \label{fig:Mr}}
\end{figure}

We also found that the colour gradients are more closely correlated with
$r_{\rm e}$ than $M_{\rm star}$, and galaxies with smaller sizes tend to show 
flatter or more positive gradients (Figs. \ref{fig:ihr} and \ref{fig:ihm}).
In Fig \ref{fig:Mr}, we show $r_{\rm e}$ vs. $M_{\rm star}$ diagram for PSBs, QGs, and SFGs with high and low $C_A$ values from the original sample of \cite{him23}.
Those PSBs with high $C_A$ values show the smallest sizes at a given $M_{\rm star}$ in galaxies with $M_{\rm star} > 10^{10} M_\odot$, and their median $r_{\rm e}$ is significantly smaller than those of PSBs with low $C_A$ and SFGs. 
Several previous studies reported small sizes of PSBs at low and intermediate redshifts (e.g., \citealp{mal18}; \citealp{wu18}; \citealp{che22}; \citealp{set22}), which could be caused by those PSBs with high $C_A$ values. 
These results may support the merger scenario for those PSBs with 
high $C_{A}$, because the intense nuclear starburst could add a significant
stellar mass at the centre and the major merger is expected to move
existing stars to inner regions, both of which lead to a decrease in size
of the galaxy (e.g., \citealp{wu20}).
We estimated the possible effect of the nuclear starburst on their sizes
by assuming an extreme case that all stars formed in the burst are added
to the centre.
The mass fraction of stars formed in the 321-1000 Myr before observation in
PSBs with $\log{C_{A}} > 0.6$ is 0.24--0.53 (16 and 84 percentiles).
We adopted these values as the burst mass fraction, and added these fractions
of mass at the centre of an exponential profile with a certain size.
We then calculated half-mass radii before and after adding the burst mass
at the centre.
The half-mass radius decreases after adding the burst mass by a factor of
1.3--1.8.
Since the colour gradients of PSBs with high $C_A$ suggest that starbursts are spatially extended to some extent, 
the effect of the starburst could be smaller in more realistic situations.
On the other hand, the weighted mean $r_{\rm e}$ of those PSBs with high $C_{A}$ for a given $M_{\rm star}$
is smaller by a factor of $\sim 2.0$ and $\sim 2.6$ than SFGs with high 
$C_{A}$ and all SFGs, respectively.
Thus additional mechanism(s) such as the major merger seems to be needed to explain the size distribution of PSBs with $\log C_A > 0.6$, while sizes of some of PSBs with high $C_A$ could be reduced by the effect of the nuclear starburst.

While normal SFGs tend to have an extended star-forming disk and
a central bulge with little star formation
(e.g., \citealp{rob94}; \citealp{blu19}), which causes their relatively
low $C_{A}$ values and negative colour gradients,
SFGs with high $C_{A}$ have significant asymmetric features
and the blue mean corrected $I-H$ colour in their central region 
(Fig. \ref{fig:cg}).
Such a blue colour suggests that active star formation occurs at their centre,
and their sizes for a given $M_{\rm star}$ are systematically smaller than
SFGs with low $C_{A}$ (Fig. \ref{fig:Mr}).
Thus the galaxy mergers/interactions may trigger gas infall into the centre
and their nuclear starburst.
Their high $C_{A}$, flat colour gradients, and relatively small sizes indicate 
that those SFGs with high $C_{A}$ could be progenitors of PSBs with high $C_{A}$.
In order to evolve into PSBs with high $C_{A}$, those SFGs with high $C_{A}$
need to quench star formation over the entire galaxy, especially,
in outer regions, and become more compact on average.
On the other hand, their mean $I-H$ colours in outer regions are slightly redder
than those of SFGs with low $C_{A}$, and therefore
those galaxies may have already begun to reduce star formation at their outskirt.

QGs with high $C_{A}$ values have similarly small (or slightly larger)
$r_{\rm e}$ with PSBs with high $C_{A}$.
Their relatively flat colour gradients are not much different from 
those of PSBs with high $C_{A}$, while their $I-H$ colours are redder
over the galaxy.
These properties of QGs with high $C_{A}$ suggest that they are descendants
of PSBs with high $C_{A}$ values.
If the relatively blue colour in the central region of PSBs with high $C_{A}$
is caused by the intense nuclear starburst in the recent past, 
their central region is expected to become redder and fainter more rapidly 
than the outer regions as time elapses, which leads to flatter 
colour gradients and slightly larger sizes.

PSBs with low $C_A$ have different physical properties from those with high $C_A$, namely, the negative mean colour gradients, larger dust extinction, and larger $r_{\rm e}$. 
Their weighted mean $r_{\rm e}$ for a given $M_{\rm star}$ is larger than those with high $C_A$ by a factor of $\sim 1.5$ and intermediate between normal SFGs and QGs (Fig. \ref{fig:Mr}). 
In Fig. \ref{fig:ihav}, we show $\Delta (I-H)$ vs. $A_V$ diagram for PSBs with low and high $C_A$. 
$\Delta(I-H)$ values tend to increase with increasing $A_V$. 
Therefore relatively high $\Delta(I-H)$ values (i.e., negative colour gradients) of PSBs with low $C_A$ are related with the dust extinction. 
For example, if there was a nuclear starburst, the dust associated with the starburst in the central region could enhance their $\Delta(I-H)$. 
On the other hand, PSBs with low $C_A$ have similar stellar mass and $C$ (Fig. \ref{fig:size}). 
When considering the PSB probability (Fig. \ref{fig:psbfr}), their weighted mean of $C$ is $\sim 5.0$, which is slightly smaller than that of PSBs with high $C_A$ ($\sim 5.4$) but still consistent with normal QGs. 

In order to consider possible effects of the centrally concentrated dust extinction, we examined the possibility that PSBs with low $C_A$ have the same stellar structure as those with high $C_A$ but young stars formed in the recent starburst are obscured by dust in these galaxies. 
For this purpose, we made a toy model by using SFHs estimated in the SED fitting, surface brightness profile, namely, $r_{20}$, $r_{\rm e}(=r_{50})$, and $r_{80}$, and the colour gradients of PSBs with high $C_A$ values.
We divided stars in a model galaxy into old population with age > 1Gyr and young one with age < 1Gyr, and calculated intrinsic $I-H$ colours of these two populations from the estimated SFHs.
With the colours, we determined mass fraction of young population at each radius so that the observed colour gradients of those PSBs with high $C_A$ are reproduced. We assumed S{\'e}rsic surface brightness profiles and no age/metallicity radial gradient for underlying old population. We fitted $r_{\rm e}$ and S{\'e}rsic index n for the old population so that total surface brightness of the old and young populations have the same $r_{20}$, $r_{\rm e}$, and $r_{80}$ as the observed PSBs with high $C_A$.
We then added the dust extinction only to the young population and recalculated $r_{\rm e}$ and $C$.
As a result, the median $r_{\rm e}$ increases by a factor of $\sim 1.2$ when dust extinction of $A_V = 2.4$ (approximately the maximum $A_V$ value for PSBs with low $C_A$) or complete obscuration (i.e., removing the young population) is applied.
The changes are insufficient to reproduce the observed $r_{\rm e}$ of PSBs with low $C_A$.
Similarly, the median value of $C$ increases by $\sim 0.14$ assuming $A_V = 2.4$ or complete obscuration.
The changes in $C$ are also insufficient for the observed $C$ values of PSBs with low $C_A$.
Thus, only the dust extinction effect is difficult to explain the differences between PSBs with low and high $C_A$. 
The mass fraction of stars formed in the burst (321--1000Myr before observation) for those PSBs with low $C_A$ is 0.22--0.40, which is smaller than those with high $C_A$. 
The weaker burst strength and their intermediate sizes between SFGs and QGs suggest that different mechanism(s) from the major merger could lead to those PSBs with low $C_A$. 
For example, minor merger, interaction with other objects, or bar instability could cause a weaker starburst in the centre and some gas and dust may not have completely lost from these galaxies after the burst (e.g., \citealp{dav19}). 
\cite{paw18} suggested that a minor merger between a massive QG and a low-mass SFG is another possible origin of massive PSBs. 
Those PSBs with low $C_A$ tend to have slightly higher Asymmetry index $A$ (Fig. 9 of \citealp{him23}). 
Such asymmetric features may be disturbances associated with such processes.
Thus, PSBs with high and low $C_{A}$ values may be caused by the different mechanisms and follow the different evolutionary paths.

\section{Summary}
We measured the colour gradients of PSBs at $0.7 < z < 0.9$ in the COSMOS field
by using the $HST$/ACS $I_{\rm F814W}$-band and $HST$/WFC3 $H_{\rm F160W}$-band
data, and compared them with those of QGs and SFGs.
Our main results are summarised as follows.

\begin{itemize}

\item The median $\Delta (I-H)$ decreases with increasing $C_A$ for
  all the three populations. PSBs and SFGs show relatively strong correlations
  between $\Delta (I-H)$ and $C_{A}$, while $\Delta (I-H)$ of QGs only weakly
  depends on $C_{A}$.

\item  PSBs with $\log{C_{A}} > 0.6$ show the lowest 
  median value of $\Delta (I-H) \sim -0.09$, which means that the colour become
  bluer with decreasing radius. The positive colour gradients and relatively
  blue $I-H$ colours in the central regions of those PSBs 
  suggest that their high $C_{A}$ values are caused by disturbed distribution
  of relatively young stellar population at the centre.

\item The colour gradients for all the three populations are more closely
  related with $r_{\rm e}$ than $M_{\rm star}$. Those galaxies with smaller sizes
  tend to show flatter or more positive colour gradients (negative $\Delta (I-H)$).
  
\item PSBs with $\log{C_{A}} > 0.6$ have the smallest half-light radii
  for a given $M_{\rm star}$, which are similar with 
  or slightly smaller than QGs with high $C_{A}$.
  On the other hand, the sizes of PSBs with $\log{C_{A}} < 0.6$ are
  larger than those with high $C_{A}$ by a factor of  $\sim 1.5$, and  
  intermediate between normal SFGs and QGs.

These results suggest that those PSBs with $\log{C_{A}} > 0.6$  
experienced a nuclear starburst in the recent past, and relatively young
stars formed in the burst cause their blue colour at the centre. 
The gas-rich major mergers may trigger the gas infall to their centre and
lead to the rapid quenching of star formation.
On the other hand, similarly massive PSBs with $\log C_A < 0.6$ show the negative colour gradients, heavier dust extinction, and larger sizes, 
and the physical origins could be different between those PSBs with
high and low $C_{A}$.

\end{itemize}

 
\section*{Acknowledgements}

We thank the referee for valuable suggestions and comments. 
This research is based in part on data collected at Subaru Telescope,
which is operated by the National Astronomical Observatory of Japan.
We are honoured and grateful for the opportunity of observing the 
Universe from Maunakea, which has the cultural, historical and natural 
significance in Hawaii.
Data analysis were in part carried out on common use data analysis computer
system at the Astronomy Data Center, ADC, of the National Astronomical
Observatory of Japan.

\section*{Data Availability}

The COSMOS2020 catalogue is publicly available at
https://cosmos2020.calet.org/.
The COSMOS {\it HST}/ACS $I_{\rm F814W}$-band mosaic data version 2.0
are publicly available via NASA/IPAC Infrared Science Archive at
https://irsa.ipac.caltech.edu/data/COSMOS/images/acs\_mosaic\_2.0/.
The COSMOS-DASH {\it HST}/WFC3 $H_{\rm F160W}$-band mosaic data version 1.2.10
are publicly available via Mikulski Archive for Space Telescopes at 
https://archive.stsci.edu/hlsp/cosmos-dash/.
The raw data for the ACS and WFC3 mosaics are also available at
https://archive.stsci.edu/missions-and-data/hst.
The Subaru/Suprime-Cam $i{'}$-band mosaic reduced data
are also publicly available 
at https://irsa.ipac.caltech.edu/data/COSMOS/images/subaru/.
The raw data for the Suprime-Cam mosaic are accessible through
Subaru Telescope Archive System at 
https://stars.naoj.org.







\appendix

\section{PSB probability}
 \begin{figure*}[h]
  \includegraphics[width=2\columnwidth]{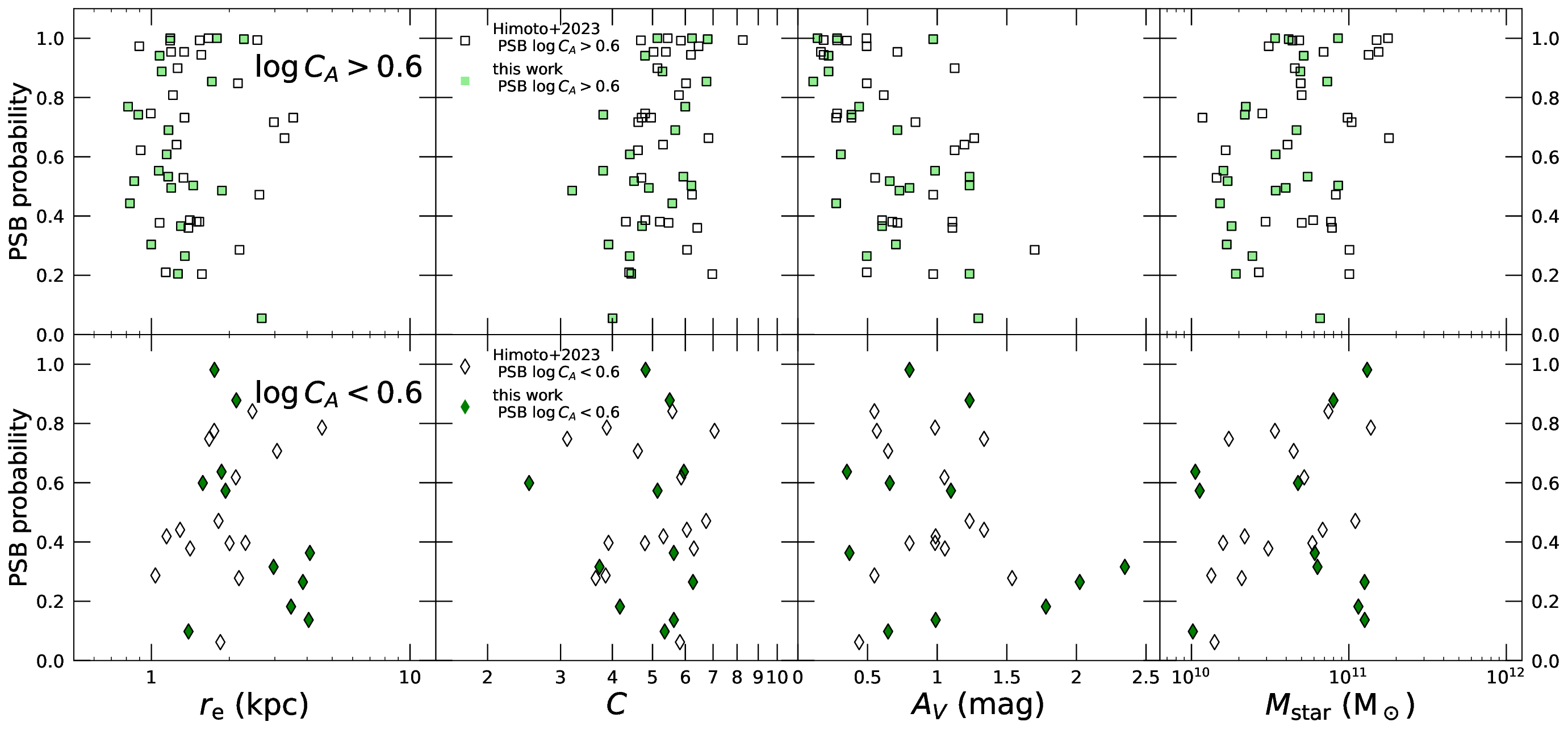}
  \caption{PSB probability as a function of $r_{\rm e}$, $C$, $A_V$, $M_{\rm star}$ for PSBs with 
  $\log C_A >0.6$ (top panels) and $\log C_A <0.6$ (bottom panels) in \citealp{him23} (open symbol) and in this study (colour symbol).
  The galaxies from \citet{him23} are limited to those with $M_{\rm star} > 10^{10} M_{\odot}$ for comparison.
    \label{fig:psbfr}}
\end{figure*}

\begin{figure}
  \includegraphics[width=\columnwidth]{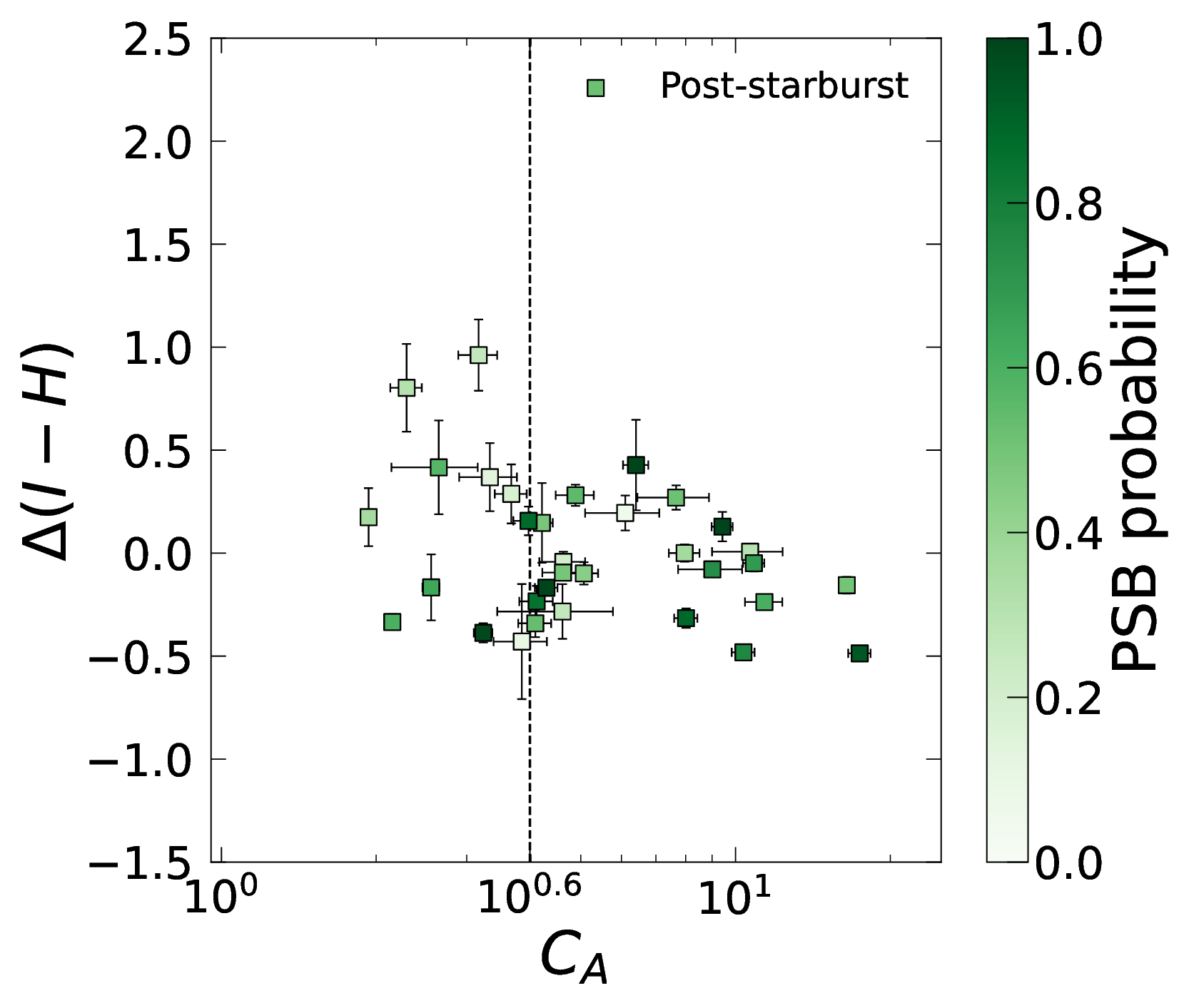}
  \caption{The same as the left panel of Fig. \ref{fig:deca}, 
  but the colour of symbols represents the PSB probability. 
    \label{fig:ih_psbfrac}}
\end{figure}

We here show distribution of the PSB probability and its relationship with physical properties for PSBs. 
We performed 1000 Monte Carlo simulations for a galaxy in the SED fitting (Section \ref{sec:sample}), 
and defined the PSB probability as a fraction of the simulations where the PSB selection criteria (equation (\ref{eq:psbsel})) were satisfied. 
The probability of unity means that the selection criteria were satisfied in all 1000 simulations.
Note that SSFR$_{\rm 0-40Myr}$, SSFR$_{\rm 40-321Myr}$, and SSFR$_{\rm 321-1000Myr}$
  used in the selection criteria are median values in the Monte Carlo
  simulations, and therefore there are PSBs with a relatively low probability
  that all the three criteria are satisfied. Such objects do not satisfy 
  any of the criteria in a relatively large fraction of the simulations,
  but still tend to have rapidly declining SFHs (Figure 6 of \citealp{him23}).
Fig. \ref{fig:psbfr} shows the PSB probability as a function of physical properties, namely,  
$r_{\rm e}$, $C$, $A_V$, and $M_{\rm star}$ for PSBs with $\log C_A >0.6$ and $\log C_A <0.6$ from \citealp{him23} and in this study.
In both the original sample from \cite{him23} and our sample, the fraction of galaxies with PSB probability less than 0.4 is slightly higher in PSBs with low $C_A$ than those with high $C_A$. 
While the PSB probability is not significantly correlated with $r_{\rm e}$, $C$, and $M_{\rm star}$ in the original sample, dusty galaxies with $A_V \gtrsim 1.5$ mag systematically have small PSB probabilities, because the uncertainty in the estimated SFHs tend to be large for such dusty red SEDs. 
Furthermore, large or massive PSBs with low $C_A$ in our sample tend to have low PSB probabilities, in particular, at $r_{\rm e} \gtrsim 3$ kpc. 
These galaxies could be strongly affected by contamination from non-PSB galaxies.
There is no such strong bias for PSBs with high $C_A$ in our sample, while PSBs with the largest $r_{\rm e}$ have a very low PSB probability.

In Fig. \ref{fig:ih_psbfrac}, we show the $\Delta (I-H)$ vs. $C_A$ diagram for PSBs, with the PSB probability.
PSBs with high $C_A$ and low $\Delta(I-H)$ exhibit relatively high PSB probabilities, which ensures that those PSBs with high $C_A$ 
preferentially show the positive colour gradients. 
On the other hand, PSBs with low $C_A$ and high $\Delta (I-H)$ tend to have relatively low PSB probabilities. 
This affects the weighted mean colour profile for PSBs with low $C_A$ values in the right panel of Fig. \ref{fig:cg}.

\section{Colour gradient with respect to physical scale} \label{sec:dlogr}
\begin{figure*} 
  \includegraphics[width=1.9\columnwidth]{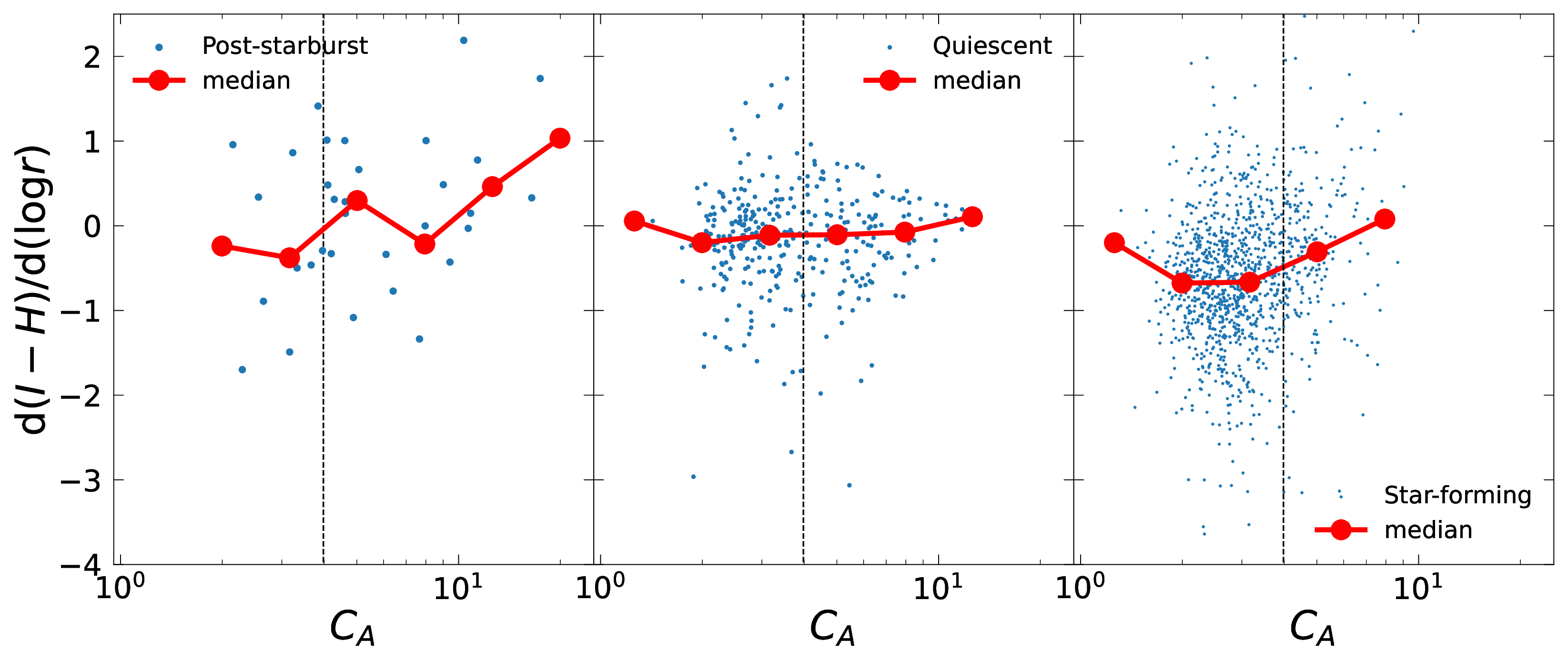}
  \caption{ 
  Colour gradient with respect to physical scale, i.e., $d(I-H)/d(\log{r})$ as a function of $C_A$ for PSBs (left), QGs (middle), and SFGs (right). 
  The inner and outer colours are measured with the same manner as in
  Fig. \ref{fig:deca}, and then $(I-H)_{\rm out} - (I-H)_{\rm in}$ are scaled by 
  $d(\log{r}) = \log{r_{\rm out}} - \log{r_{\rm in}}$, where $r_{\rm in}$ and
  $r_{\rm out}$ are in unit of kpc.
  Circles show the median values of the colour gradient
  in $C_A$ bins with a width of $\pm 0.1$ dex.
  \label{fig:cgrca}}
\end{figure*}

\begin{figure} 
  \includegraphics[width=\columnwidth]{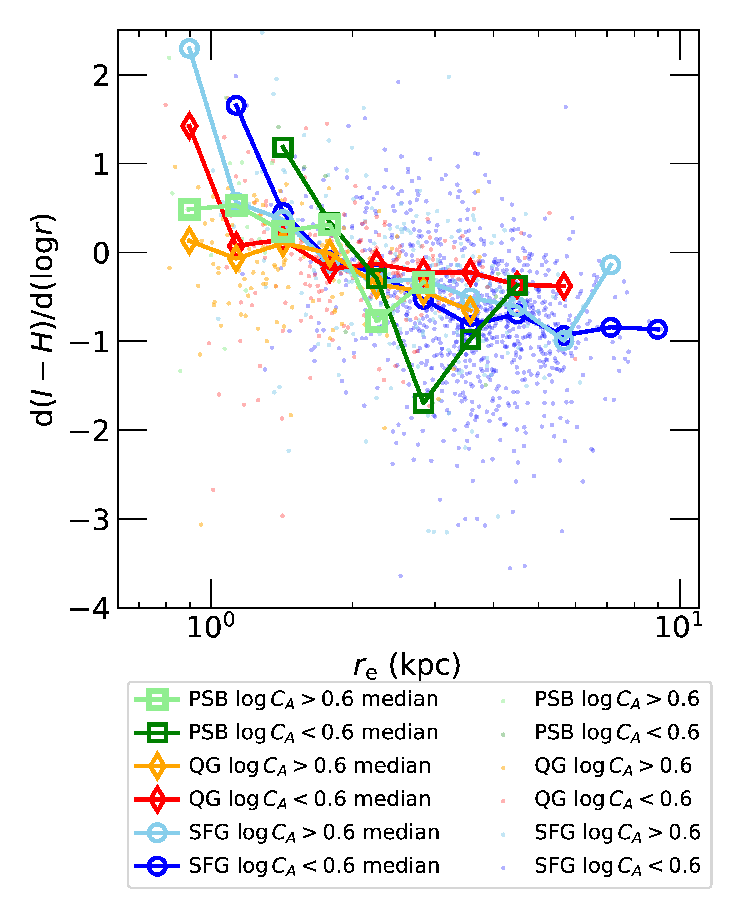}
  \caption{ 
    $d(I-H)/d(\log{r})$ vs. $r_{\rm e}$ for PSBs, QGs, and SFGs
    with $\log C_A <0.6$ and $\log C_A>0.6$. 
  The symbols are the same as Fig. \ref{fig:ihr}.
  \label{fig:cgrre}}
\end{figure}

  We here present the colour gradient with respect to physical scale, namely, 
$d(I-H)/d(\log{r})$. Fig. \ref{fig:cgrca} shows $d(I-H)/d(\log{r})$ as a
function of $C_{A}$ for PSBs, QGs, and SFGs.
While there is a relatively large scatter, the median $d(I-H)/d(\log{r})$
for PSBs increases with increasing $C_{A}$, and PSBs with high $C_{A}$ values 
tend to have positive colour gradients. This result is consistent with that 
in Fig. \ref{fig:deca}, where we scaled the colour gradient with Kron radius
of the object.
The colour gradients of QGs distribute around $d(I-H)/d(\log{r}) \sim 0$,
and do not strongly depend on $C_{A}$, although their median value slightly
increases with $C_{A}$.
Most SFGs tent to have relatively low $C_{A}$ and negative colour gradients,
while those with high $C_{A}$ show flatter gradients.
Their median $d(I-H)/d(\log{r})$ increases with increasing $C_{A}$.

Fig. \ref{fig:cgrre} shows $d(I-H)/d(\log{r})$ vs. $r_{\rm e}$ for PSBs, QGs, and
SFGs with low and high $C_{A}$ values.
The median $d(I-H)/d(\log{r})$ decreases with increasing $r_{\rm e}$ for all
the populations, and those galaxies with small $r_{\rm e}$ tend to have positive
colour gradients. These trends are consistent with Fig. \ref{fig:ihr}.
Thus our results do not strongly change if we use $d(I-H)/d(\log{r})$
instead of $\Delta(I-H)$.

\section{Colour gradient as a function of $A$ and $C$}\label{sec:asycon}

\begin{figure*} 
  \includegraphics[width=1.9\columnwidth]{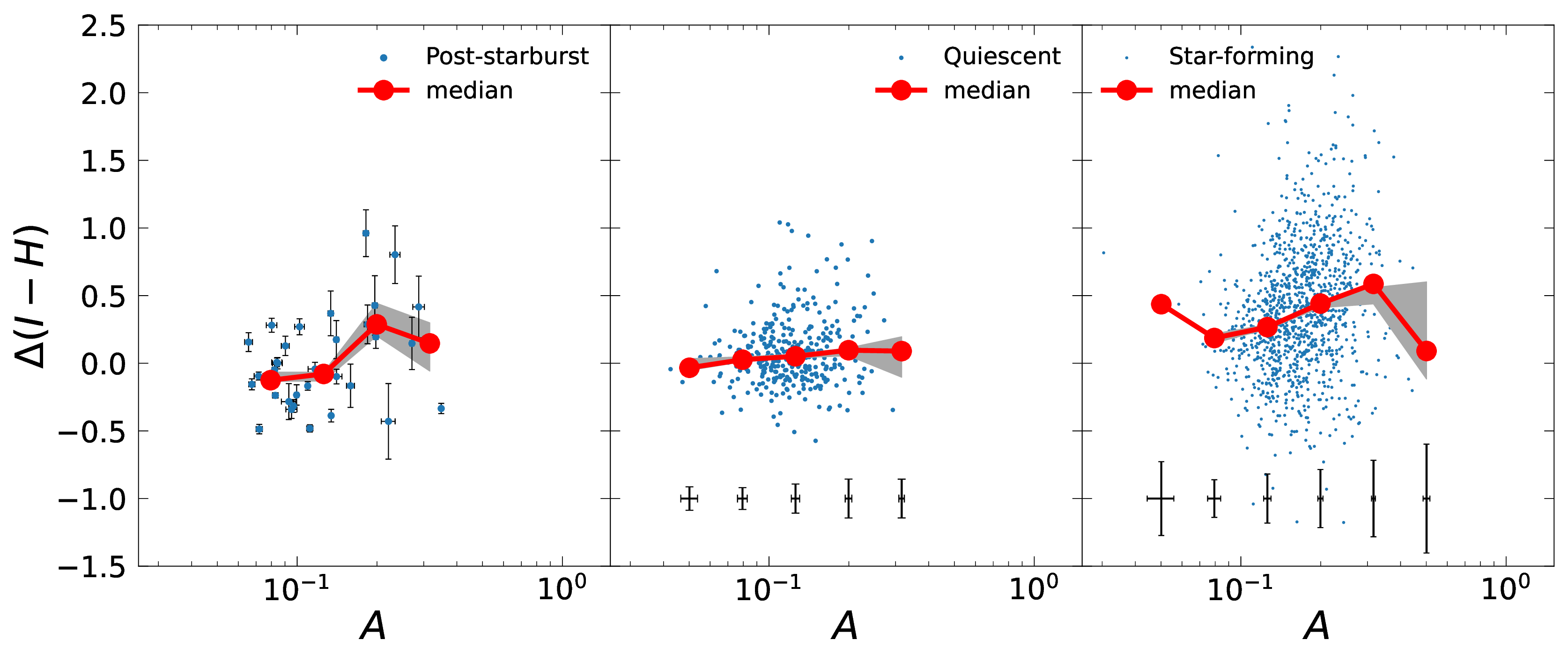}
  \caption{ 
    $\Delta(I-H)$ as a function of $A$ for PSBs (left), QGs (middle),
    and SFGs (right). 
  The symbols are the same as Fig. \ref{fig:deca}.
  \label{fig:deasy}}
\end{figure*}

\begin{figure*} 
  \includegraphics[width=1.9\columnwidth]{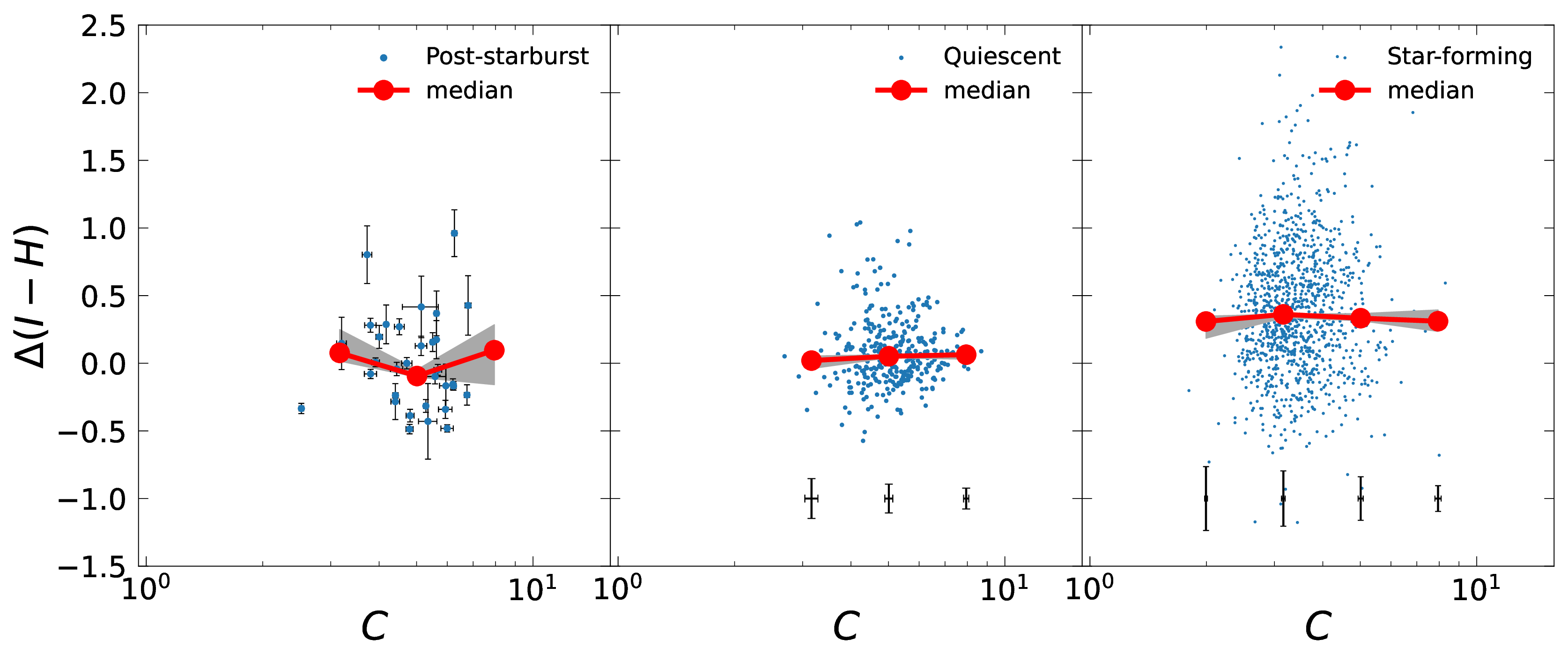}
  \caption{ 
    $\Delta(I-H)$ as a function of $C$ for PSBs (left), QGs (middle),
    and SFGs (right). 
  The symbols are the same as Fig. \ref{fig:deca}.
  \label{fig:decon}}
\end{figure*}

We here present the colour gradients for our sample galaxies 
  as a function of Asymmetry index $A$ and Concentration index $C$. 
  Figs. \ref{fig:deasy} and \ref{fig:decon} show $\Delta(I-H)$ vs. $A$,
  and $\Delta(I-H)$ vs. $C$, respectively, for PSBs, QGs, and SFGs.
In Fig. \ref{fig:deasy}, PSBs with $A < 0.15$ tend to have
$\Delta I-H \lesssim 0.3$, and their median $\Delta I-H$ is lower than
that of PSBs with $A \gtrsim 0.15$.
Since most of PSBs with $A < 0.15$ have $\log{C_{A}}>0.6$
(Fig. \ref{fig:caac}), this trend is consistent with the $C_{A}$ dependence
of $\Delta I-H$ seen in Fig. \ref{fig:deca} and Table \ref{tab:example_table}.
The colour gradients of QGs do not strongly depend on $A$, although the median
$\Delta I-H$ slightly increases with increasing $A$.
The median $\Delta I-H$ of SFGs clearly increases with increasing $A$ at
$A \sim$ 0.08--0.3, while there is a relatively large scatter at a given
$A$.
On the other hand, one can see that $\Delta I-H$ does not significantly
depend on $C$ for all three populations in Fig. \ref{fig:decon}.
These results and Fig. \ref{fig:deca} 
suggest that the asymmetric features, in particular, $C_{A}$
are closely related with the colour gradients of galaxies.



\bsp	
\label{lastpage}
\end{document}